\documentclass[manuscript]{aastex}

\usepackage{subfigure}
\usepackage{natbib}
\usepackage{lscape}

\newcommand{\rmn}[1]{{\mathrm{#1}}}
\newcommand{\kw}{Konus-\textit{Wind}\ }
\newcommand{\kws}{KW\ }
\newcommand{\ntot}{1939\ }

\shorttitle{The second \kws catalog of short GRBs}
\shortauthors{Svinkin et al.}

\begin{document}


\title{\bfseries The second \kw catalog of short gamma-ray bursts}


\author{D.~S.~Svinkin, D.~D.~Frederiks, R.~L.~Aptekar, S.~V.~Golenetskii,
        V.~D.~Pal'shin, Ph.~P.~Oleynik, A.~E.~Tsvetkova, M.~V.~Ulanov,}
\affil{Ioffe Institute, Politekhnicheskaya 26, St.~Petersburg, 194021, Russia}

\author{T.~L.~Cline\altaffilmark{1},}
\affil{NASA Goddard Space Flight Center, Greenbelt, MD 20771, USA}
\altaffiltext{1}{Emeritus}
\and
\author{K.~Hurley}
\affil{Space Sciences Laboratory, University of California,
7 Gauss Way, Berkeley, CA 94720-7450, USA}




\begin{abstract}
In this catalog, we present the results of a systematic study of 295 short
gamma-ray bursts (GRBs) detected by \kw (KW) from 1994 to 2010.
From the temporal and spectral analyses of the sample, we provide the burst
durations, the spectral lags, the results of spectral fits with three model functions,
the total energy fluences and the peak energy fluxes of the bursts.
We discuss evidence found for an additional power-law spectral component and the presence of extended
emission in a fraction of the KW short GRBs.
Finally, we consider the results obtained in the context
of the Type I (merger-origin) / Type II (collapsar-origin) classifications.
\end{abstract}

\keywords{gamma-ray burst: general~--- catalogs}

\section{INTRODUCTION}
Gamma-ray bursts (GRBs) can be divided into two distinct
morphological classes based on the properties of the observed
gamma-ray emission: short/hard GRBs, which typically last less than 2~s,
have hard prompt-emission spectra and negligible spectral lag,
and long/soft GRBs which last typically longer than 2~s, have softer spectra
and non-negligible spectral lag~\citep{Mazets_1981_part_1,Kouveliotou_1993ApJ,Norris_2000ApJ,Norris_and_Bonnel_2006ApJ}.

It is believed that the physical origins of long/soft and short/hard bursts are different.
Short/hard GRBs are thought to be the results of mergers of binary compact objects (so called Type~I GRBs),
such as two neutron stars or a neutron star and a black hole (see, e.g.~\citealp{Berger_2014ARAA}
and references therein), while long/soft (Type~II GRBs),
which are occasionally accompanied by supernovae, originate from the core collapse of massive stars
(see~\citealp{Zhang_2009ApJ} for more information on the Type I/II classification scheme).

The \kw gamma-ray burst spectrometer (hereafter KW, \citealp{Aptekar_1995SSR}) has observed $\sim 2500$ GRBs,
with $\sim 400$ of them being short GRBs, in the period from launch in 1994 to 2015.
Here, we present the second \kws short GRB catalog which provides spectral and temporal
characteristics of one of the largest short GRB samples to date over a broad energy band.
Specifically, the catalog covers GRBs occurring during the period from 1994 November to 2010 December
and includes about twice the number of bursts as the first Konus catalog
of short GRBs\footnote{The data are available at http://www.ioffe.ru/LEA/shortGRBs/Catalog/}~\citep{Mazets_2002astroph}.

We start with a description of the \kws detectors in Section~\ref{sec:KW}.
In Section~\ref{sec:OBSERVATIONS} we provide details of the \kws short GRB sample.
We describe the analysis procedures in Section~\ref{sec:DATA ANALYSIS} and
present the results in Section~\ref{sec:RESULTS}.
Finally, in Section~\ref{sec:SUMMARY} we conclude with a summary.

\section{KONUS-WIND}\label{sec:KW}
\kws consists of two identical NaI(Tl) detectors S1 and S2, each with $2\pi$ field of view.
The detectors are mounted on opposite faces of the rotationally stabilized
\textit{Wind} spacecraft, such that one detector~(S1) points towards the south ecliptic pole,
thereby observing the south ecliptic hemisphere, while the other~(S2) observes the north ecliptic hemisphere.
Each detector has an effective area of $\sim 80$--160~cm$^2$ depending on
the photon energy and incident angle. The nominal energy range of gamma-ray measurements covers
the incident photon energy interval from 13~keV up to 10~MeV.

In interplanetary space far outside the Earth's magnetosphere, \kws has the advantages
over Earth-orbiting GRB monitors of continuous coverage, uninterrupted by Earth occultation,
and a steady background, undistorted by passages through the Earth's trapped radiation,
and subject only to occasional solar particle events.
The \textit{Wind} distance from Earth as a function of time is presented in~\cite{Palshin_2013ApJS}.
The maximum distance was $\sim 7$~lt-s in 2002 January and May;
since 2004 \textit{Wind} has been in a Lissajous orbit at
the $L_1$ libration point of the Sun-Earth system at a distance of $\sim 5$~lt-s.

The instrument has two operational modes: waiting and triggered. While in the waiting
mode, the count rates are recorded in three energy windows G1~(13--50~keV), G2~(50--200~keV),
and G3~(200--760~keV) with 2.944 s time resolution.
When the count rate in the G2 window exceeds a $\approx 9\sigma$ threshold
above the background on one of two fixed time-scales, 1~s or 140~ms, the instrument switches into the triggered mode. 
In the triggered mode, the count rates in the three energy windows are recorded
with time resolution varying from 2~ms up to 256~ms. These time histories, with a total
duration of $\sim 230$~s, also include 0.512~s of pre-trigger history.
Spectral measurements are carried out, starting from the trigger time $T_0$,
in two overlapping energy intervals, 13--760~keV and 160~keV--10~MeV,
with 64 spectra being recorded for each interval over a 63-channel, pseudo-logarithmic energy scale.
The first four spectra are measured with a fixed accumulation time of 64~ms
in order to study short bursts. For the subsequent 52 spectra, an adaptive system
determines the accumulation times, which may vary from 0.256 to 8.192~s depending
on the current count rate in the G2~window. The last 8 spectra are obtained for 8.192~s each.
As a result the minimum duration of spectral measurements is 79.104~s, and the maximum is 491.776~s.

The detector response matrix (DRM), which is a function only of the energy and incident angle,
was computed using the GEANT4 package~\citep{Agostinelli_2003NIMPA}.
The detailed description of the instrument response calculation is presented in~\citet{Terekhov_1998AIPC}.
The latest version of the DRM contains responses calculated for 255 photon energies between 5~keV
and 26~MeV on a quasi-logarithmic scale for incident angles from $0\degr$
to $90\degr$ with a step of $5\degr$. The energy scale is calibrated in-flight using
the 1460~keV line of $^{40}$K and the 511~keV annihilation line.

The gain of the detectors has slowly decreased during the long period of operation.
The instrumental control of the gain became non-functional in 1997 and the spectral range changed
to 25~keV--18~MeV for the S1 detector and to 20~keV--15~MeV for the S2 detector,
from the original 13~keV--10~MeV.
The spectral resolution of the detectors ($\Delta E/E$) did not change significantly during the mission,
with an upper limit, estimated at $E=1460$~keV, of $\Delta E/E \lesssim 10$\% (FWHM) 
for the whole period of monitoring.
The corresponding resolution loss is less than a factor of 1.5 compared
to the ground-based calibrations ($\Delta E/E \approx 6.5$\% at 1460~keV, FWHM).

For all short GRBs we use a standard KW dead time (DT) correction procedure for light curves
(with a DT of a few microseconds) and spectra (with a DT of $\sim$42 microseconds).
Although the photon flux for some short GRBs is very high (up to $\sim 10^5$~counts~s$^{-1}$),
this procedure is still applicable; no additional correction, which was used,
e.g., in an analysis of the KW detection of the 1998 August 27 giant flare from
SGR~1900+14, is required (details of these simulations and the
KW dead-time correction procedures can be found in \citealt{Mazets1999AstL}).
Also, at high count rates, a pile-up effect in the analog electronics
can distort the low-energy part of the KW instrumental spectra.
Our simulations show that, for the bright, hard bursts in our sample,
the distorted energy range is limited to $<$50--150 keV and lies well below
the peak energies of the spectra. Accordingly, we found that the exclusion of the
potentially distorted channels from spectral fits of the brightest short GRBs
results in model parameter changes within the fit uncertainties.

The consistency of the KW spectral parameters with those obtained in other GRB experiments
was verified by a cross-calibration with \textit{Swift}-BAT and \textit{Suzaku}-WAM~\citep{Sakamoto_2011PASJ},
and in joint spectral fits with \textit{Swift}-BAT~(\citealt{Krimm_2006ApJ,Roming_2006ApJ,Starling_2009MNRAS})
and \textit{Fermi}-GBM~(e.g., \citealt{Lipunov_2016MNRAS}).
It was shown that the difference in the spectrum normalization between KW and
these instruments is $\lesssim 20$\% in joint fits.

\section{THE SHORT GRB SAMPLE}\label{sec:OBSERVATIONS}
Between 1994 November and 2010 December, \kws detected \ntot GRBs in the triggered mode,
295 of which were classified as short-duration GRBs or short bursts with extended emission (EE).
The classification (Svinkin~et~al.,~in~prep.) was based on the $T_{50}$ duration
distribution.
$T_{90}$ and $T_{50}$ are the time intervals which contain from 5\% to 95\% ($T_{90}$)
and from 25\% to 75\% ($T_{50}$) of the total burst count fluence (see, e.g.,~\citealp{Kouveliotou_1993ApJ}).
In this work, these durations are calculated in the G2+G3 band (nominal
bounds 50--760~keV) unless stated explicitly.
Using an unbiased sample of 1168 \kws GRBs we adopted $T_{50}=0.6$~s
as the boundary between short and long \kws GRBs.
The instrument trigger criteria cause undersampling of faint short
bursts relative to faint long bursts, so this subsample of fairly
bright (in terms of peak count rate in the \kws trigger energy band)
bursts has been chosen for the purpose of classification.

Although the aim of this work is to report on all \kws GRBs that meet the short GRB criterion,
the results of a more sophisticated classification of the selected GRBs,
which accounts for the burst spectral hardness and its duration (Svinkin~et~al.,~in~prep.),
may be essential for future analysis.
The burst spectral hardness ($\rmn{HR}_{32}$) was calculated using the ratio of counts in
the G3 and G2 bands accumulated during the burst duration $T_{100}$.
The calculation of $\rmn{HR}_{32}$ takes into account the gain drift effect.
The rates expected in the nominal G2 and G3 energy bands (as given in Section~\ref{sec:KW})
were estimated using the best fit to the burst count spectrum with the CPL function
(see Section~\ref{sec:DATA ANALYSIS} for the CPL spectral model definition).

\begin{figure}
    \begin{center}
        \includegraphics[width=0.75\textwidth]{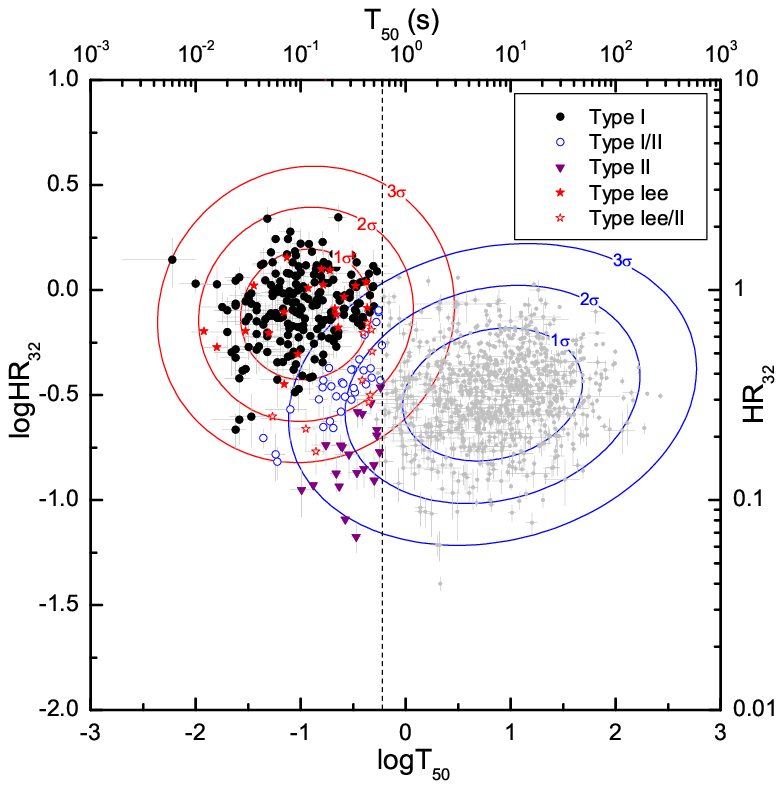}
    \end{center}
    \caption{Hardness-duration distribution of 1143 \kw bright GRBs.
    The distribution is fitted by a sum of two Gaussian distributions using
    the expectation and maximization (EM) algorithm.
    The contours denote $1\sigma$, $2\sigma$, and $3\sigma$ confidence regions
    for each Gaussian distribution.
    The vertical dashed line denotes the boundary ($T_{50}=0.6$~s) between long and short \kws GRBs.
    The types for GRBs with $T_{50}<0.6$~s are shown in colors.
    \label{fig:HRvsT50}}
\end{figure}

The burst types were derived using a method similar to that described in~\citet{Horvath_2010ApJ} and are as follows:
I (merger-origin), II (collapsar-origin), I/II (the type is uncertain).
The following correspondence between the short/hard burst indicator function
$I(T_{50},\rmn{HR}_{32})$ (see eq.~5 in~\citealp{Horvath_2010ApJ}) and the Type was used:
$I > 0.9$~--- Type~I, $0.1 < I < 0.9$~--- Type~I/II, $I < 0.1$~--- Type~II.
The classification of short GRBs with extended emission (see Section~\ref{sec:EE})
was based on the initial pulse parameters and the types are as follows:
Iee (type I which shows extended emission, EE),
and Iee/II (the type is uncertain: Iee or II).
The classification results are shown in Figure~\ref{fig:HRvsT50}.

Along with the Type I--II classification we report the spectral lags ($\tau_\rmn{lag}$) for bursts in our sample.
The spectral lag is a quantitative measure of spectral evolution often seen in long GRBs,
when the emission in a soft detector band peaks later relative to a hard band.
It was also shown that short GRBs with and without EE have negligible
spectral lag~\citep{Norris_and_Bonnel_2006ApJ, Norris_2001conf}.
Thus, the spectral lag can be used as an additional classification parameter.
We calculated $\tau_\rmn{lag}$ for three pairs of the KW light curves (G2 and G1, $\tau_\rmn{lag21}$; G3 and G2, $\tau_\rmn{lag32}$;
and G3 and G1, $\tau_\rmn{lag31}$) using a cross-correlation method similar to that described in~\citet{Norris_2000ApJ}.
Details and examples of lag estimations for KW bursts will be given in Svinkin~et~al.~(in~prep).

\kws has only coarse localization capability on its own, which is crucial for the GRB spectral analyses.
In cases where the position of a GRB is not available from an instrument with
imaging capabilities (e.g. \textit{Swift}-BAT), the source localization can
be derived using InterPlanetary network triangulation~\citep{Hurley_2013EAS}.
The localizations of 296 \kws short GRBs detected between 1994 November and 2010 December
can be found in~\citep{Palshin_2013ApJS}.

Table~\ref{tab:info} lists the 295 \kws short GRBs. The first column gives the
burst designation in the form ``GRBYYYYMMDD\_Tsssss'', where YYYYMMDD is the burst date, and
sssss is the \kws trigger time $T_0$ (UT) truncated to integer seconds
(note that due to \textit{Wind}'s large distance from Earth, this trigger time can differ
by up to $\sim 7$~s from the Earth-crossing time; see~\citealp{Palshin_2013ApJS}).
The second column gives the \kws trigger time in the standard time format.
The ``Name''\ column specifies the GRB name as provided in
the Gamma-ray Burst Coordinates Network circulars\footnote{http://gcn.gsfc.nasa.gov/gcn3\_archive.html}, if available.
The ``Detector''\ column specifies the triggered detector.
The next column provides the angle between the GRB direction and the detector axis (the incident angle).
The last column contains localization-specific notes.

Our sample contains 19 GRBs with incident angles close to, but slightly greater, than $90\degr$.
In these cases we use an incident angle of $90\degr$ to calculate the detector response.
The positions of three weak bursts, GRB19990831\_T41835, GRB20010420\_T30786, and GRB20080321\_T23721,
cannot be constrained to better than an ecliptic hemisphere,
so for these GRBs we use an incident angle of $60\degr$.

\section{SPECTRAL ANALYSIS}\label{sec:DATA ANALYSIS}

A typical \kws short GRB spectrum is a subset of the four 64-ms spectra measured from $T_0$ up to $T_0+0.256$~s.
The background spectrum for bursts without EE was usually taken from
$T_0+25$~s with an accumulation time of about 100~s.
For about 25\% of the bursts, a major fraction of the counts was accumulated before the trigger,
in the time interval not covered by the multichannel spectra.
For these bursts, a three-channel spectrum, constructed from the light curve counts,
is used for the analysis, accumulated over the whole burst duration $T_{100}$.
The spectral sample contains 214 multichannel time-integrated spectra
and 79 three-channel spectra.
Due to low counting statistics of the majority of our short GRBs we typically use a time-integrated spectrum
to calculate both the total energy fluence ($S$) and the peak energy flux~($F_\rmn{peak}$).
Only for 18 fairly intense GRBs from our sample was it possible
to derive $F_\rmn{peak}$ from a spectrum covering a narrow time interval near the peak count rate.

We chose three spectral models to fit the spectra of GRBs from our sample.
These models were a power law (PL), Band's GRB function (BAND),
and an exponential cutoff power-law (CPL).
The details of each model are presented below.

The power law model:
\begin{equation}
f_{\rmn{PL}} \propto E^{\alpha}
\end{equation}

The exponentially cutoff power law:
\begin{equation}
f_{\rmn{CPL}} \propto E^{\alpha}
                   \exp\left(-\frac{E (2+\alpha)}{E_\rmn{p}}\right)
\end{equation}

Band's GRB function~\citep{Band_1993ApJ}:
\begin{equation}
f_{\rmn{BAND}} \propto \left\{
\begin{array}{lr}
E^{\alpha} \exp\left(-\frac{E (2+\alpha)}{E_\rmn{p}}\right)
                                                    & E < (\alpha-\beta) E_\rmn{p}/(2+\alpha)  \\
E^{\beta}
      \exp(\beta-\alpha) \left[\frac{(\alpha-\beta) E_\rmn{p}}{(2+\alpha)} \right] &
    E \ge (\alpha-\beta) E_\rmn{p}/(2+\alpha)\\
\end{array}
\right.
\end{equation}
Where $E_\rmn{p}$ is the peak energy of the $E F_E$ spectrum.

\subsection{Multichannel spectra}
The spectral analysis of bursts with multichannel spectra was performed using
\texttt{XSPEC}~v.~12.8.0~\citep{Arnaud_1996ASPC}.
The $\chi^2$ statistic was used in the model fitting process as a figure of merit
to be minimized. The spectral channels were grouped to have a minimum of 10
counts per channel to ensure the validity of the $\chi^2$ statistic.
We use a model energy flux in the 10~keV--10~MeV band as the model normalization during fit.
The flux was calculated using the \texttt{cflux} convolution model in \texttt{XSPEC}.     
The parameter errors were estimated using the \texttt{XSPEC} command \texttt{error}
based on the change in fit statistic ($\Delta \chi^2 = 2.706$) which corresponds to 90\%~CL.

We fit the three model functions described above to each multichannel spectrum.
The most preferred model (the best-fit model) was chosen based on the difference in $\chi^2$.
The criterion for accepting a model with a single additional parameter is a change in $\chi^2$
of at least 5 with the chance probability for achieving this difference of $\approx 0.025$.
We found that this threshold is preferred, for our sample, over the frequently-used $\Delta \chi^2 \geq 6$
because it nearly halves the number of divergent PLs as best-fit models which is crucial when the burst
energetics are considered.

\subsection{Three-channel spectral analysis}
The 79 three-channel spectra were fitted with PL and CPL using a custom-built routine
and the confidence limits for the parameters were estimated via the bootstrap approach.

For the purpose of testing the procedure we compared results of the multichannel
and three-channel spectral analysis for the sample of the 214 GRBs with multichannel spectra.
For each burst we constructed a three-channel spectrum
accumulated over the interval of measurement of the multichannel spectrum.
Then we compared parameters of the model that best fits the multichannel spectrum
with the parameters of the same model fitted to the three-channel spectrum.
In the case of PL the resulting photon indices are consistent between the two types
of spectral analysis. For CPL we found that the $\alpha$ values are also generally
consistent between the three-channel and multichannel spectra. The same is true for $E_\rmn{p}$
but only when its value is located within the three-channel analysis range (i.e. $\lesssim 1$~MeV),
otherwise, the latter method results in an overestimated, poorly constrained $E_\rmn{p}$.
This demonstrates that \kws is capable of producing accurate spectral parameter and
energetics estimates even when multichannel spectral data are not available.

Since the CPL fit to a three-channel spectrum has zero degrees of freedom
(and, in the case of convergence,  ${\chi^2_\rmn{CPL} = 0}$), the best-fit model cannot be easily chosen between
PL and CPL on the basis of the $\Delta \chi^2\geq 5$ criterion.
So, in order not to overestimate the burst energetics, we decided to use the CPL
model flux to calculate $S$ and $F_\rmn{peak}$ for the GRBs for
which the three-channel CPL fit results in an $E_\rmn{p}$ constrained from below.

\section{RESULTS}\label{sec:RESULTS}

\subsection{Temporal characteristics and Type~I--II classification}

\begin{figure}
	\begin{center}
		\subfigure[]{\label{fig:Lags}\includegraphics[height=0.45\textwidth]{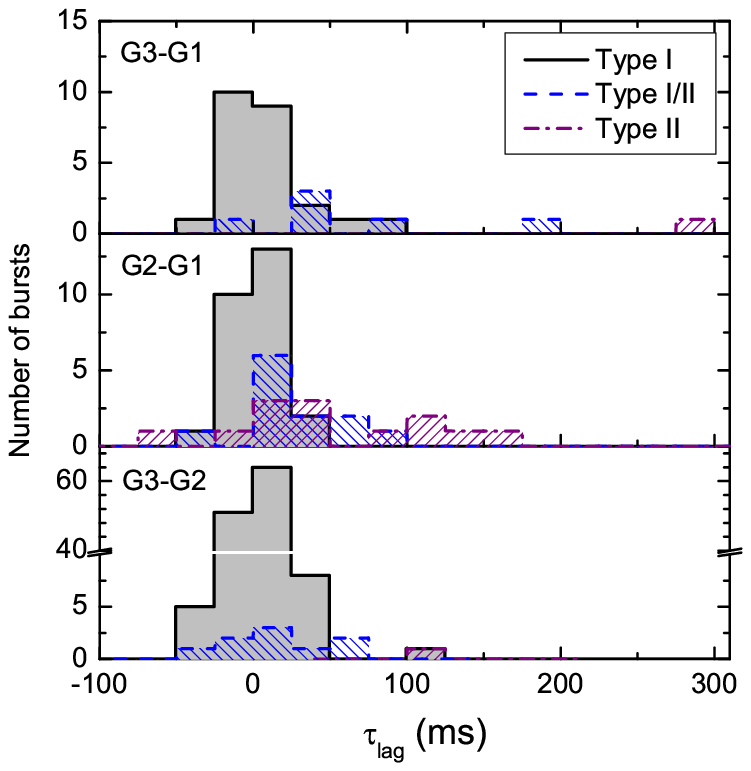}}
		\subfigure[]{\label{fig:LagsEE}\includegraphics[height=0.45\textwidth]{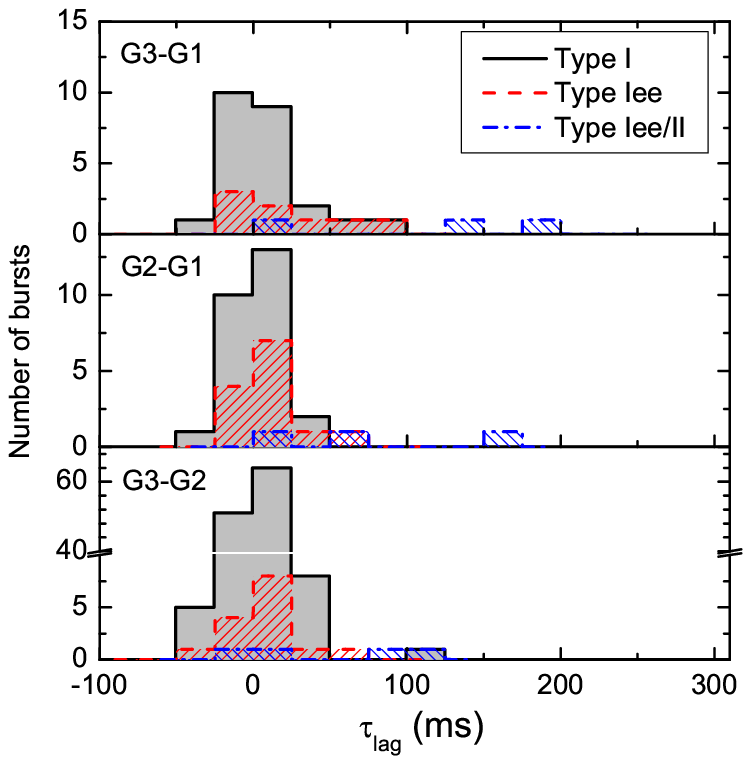}}
	\end{center}
    \caption{Spectral lag distributions of short KW GRBs without~\subref{fig:Lags} and
     with EE~\subref{fig:LagsEE}. Panel~\subref{fig:Lags} shows: Type~I bursts (gray filled histogram),
     Type~I/II bursts (dashed histogram), and Type~II bursts (dash-dotted histogram).
     Panel~\subref{fig:LagsEE} shows: Type~I bursts (gray filled histogram),
     Type~Iee bursts (dashed histogram), and Type~Iee/II bursts (dash-dotted histogram).
     \label{fig:lag_dist} }
\end{figure}

Table~\ref{tab:durations} contains the burst durations, the spectral lags, and classification.
The first column gives the burst designation.
The following four columns contain the start of the $T_{100}$ interval $t_0$
(relative to $T_0$), $T_{100}$, $T_{90}$ and $T_{50}$.
The errors are given at the $1\sigma$ confidence level~(CL).
For two GRBs (GRB19960325\_T69892 and GRB19980614\_T31854) $T_{90}$ and $T_{50}$ were
calculated in the G2 energy range (nominal bounds 50--200~keV)
because of gaps in the G3 light curve, and these GRBs were excluded from the spectral analyses.
The next column gives the Type~I--II classification and the last three columns contain
$\tau_\rmn{lag21}$,  $\tau_\rmn{lag32}$, and $\tau_\rmn{lag31}$.
A positive $\tau_\rmn{lag}$ corresponds to the delay of the softer emission.
The lags are provided for bursts having signal-to-noise ratio $\geq8$~$\sigma$
in the corresponding light curves binned to $\leq 64$~ms resolution.
Most of the Type~I bursts have $\tau_\rmn{lag}<25$~ms (see Figure~\ref{fig:lag_dist}), while bursts of
types I/II and II tend to have long $\gtrsim 100$~ms.
For the bursts with EE, the durations and lags in Table~\ref{tab:durations}
are given for the short initial pulse only.

\subsection{Spectral parameters}

Table~\ref{tab:spec_par} provides the results of the multichannel spectral analysis
for the 214 time-integrated spectra and the 18 spectra near the peak count rate.
For the time-integrated spectra the statistics of the best-fit models are as follows:
CPL~--- 201 GRBs, BAND~--- 9 GRBs, and PL~---~4 GRBs.
Along with the best-fit model parameters we present the results for the models
whose parameters are constrained (hereafter, GOOD models).
For the CPL and BAND GOOD models we require both $\alpha$ and $E_\rmn{p}$
errors to be constrained, and, for the BAND model,  $\beta > -4$.
To reject models with apparent systematics in the fit residuals,
we also require a null hypothesis probability $P> 10^{-6}$ for the fit.
The ten columns in Table~\ref{tab:spec_par} contain the following information:
(1) the burst designation (see Table~\ref{tab:info});
(2) the spectrum type, where `i' indicates that the spectrum is time-integrated and is
used to calculate $S$, `p' means that the spectrum is measured
near the peak count rate (and is used to calculate $F_\rmn{peak}$), or both `i,p';
columns (3) and (4) contain the spectrum start time $T_\rmn{start}$ (relative to $T_0$)
and its accumulation time $\Delta T$;
(5) GOOD models for each spectrum;
(6)--(8) low-energy spectral index $\alpha$, high-energy spectral index $\beta$,
and $E_\rmn{p}$;
(9) normalization (energy flux in 10~keV--10~MeV band);
(10) $\chi^2/\rmn{dof}$ along with the null hypothesis probability $P$.
In cases where the lower limit for $\beta$ is not constrained, the value of ($\beta_\rmn{min} - \beta$)
is provided instead, where $\beta_\rmn{min}=-10$ is the lower limit for the fits.
In total, the table contains results for 473 fits of time-integrated spectra with different models
(210~--- CPL, 117~--- BAND, and 146~--- PL).

Table~\ref{tab:spec_par_3ch} contains results obtained from fits of the 79 three-channel
spectra. For all but one of these GRBs we present the CPL model parameters
and, for GRB19961113\_T80522, for which $E_\rmn{p}$ is not constrained, the PL
fit results ($\chi^2 = 1.6\times 10^{-4}$) are provided.
A nonzero $\chi^2_\rmn{CPL}$ was obtained only for two out of the 78 bursts:
GRB20000623\_T03887 ($\chi^2 = 0.4$) and GRB20100612\_T47056 ($\chi^2 = 2.2$).
In both cases no excess in the count rate over the background level is detected
in the softest KW channel~(G1).
The seven columns contain the following information:
(1) the burst designation (see Table~\ref{tab:info});
columns (2) and (3) contain the spectrum start time $T_\rmn{start}$ (relative to $T_0$) 
and its accumulation time ($\Delta T$);
(4) the spectral model;
columns (5) and (6) contain $\alpha$ and $E_\rmn{p}$, respectively;
(7) normalization (energy flux in the 10~keV--10~MeV band).

\begin{figure}
	\begin{center}
		\subfigure[]{\label{fig:alpha}\includegraphics[height=0.45\textwidth]{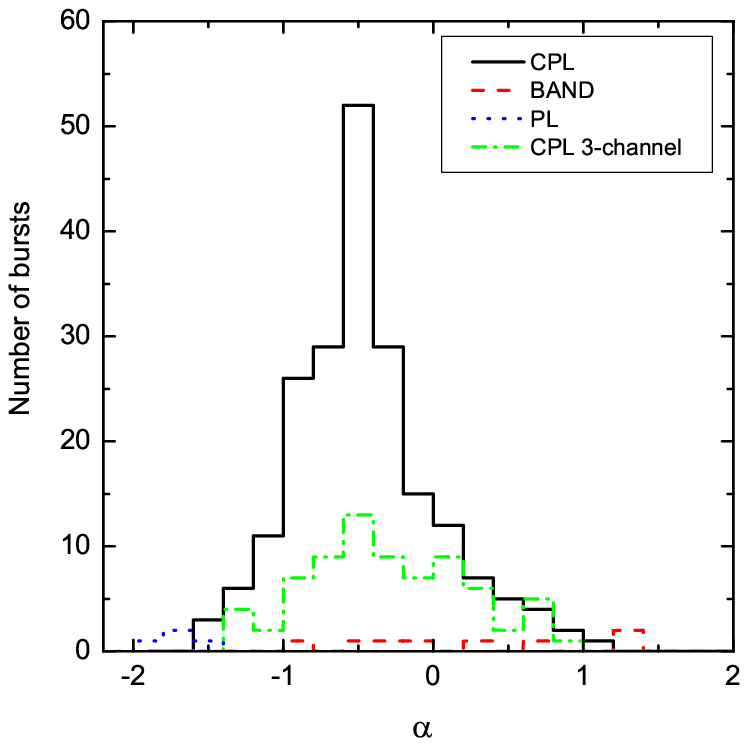}}
		\subfigure[]{\label{fig:Ep}\includegraphics[height=0.45\textwidth]{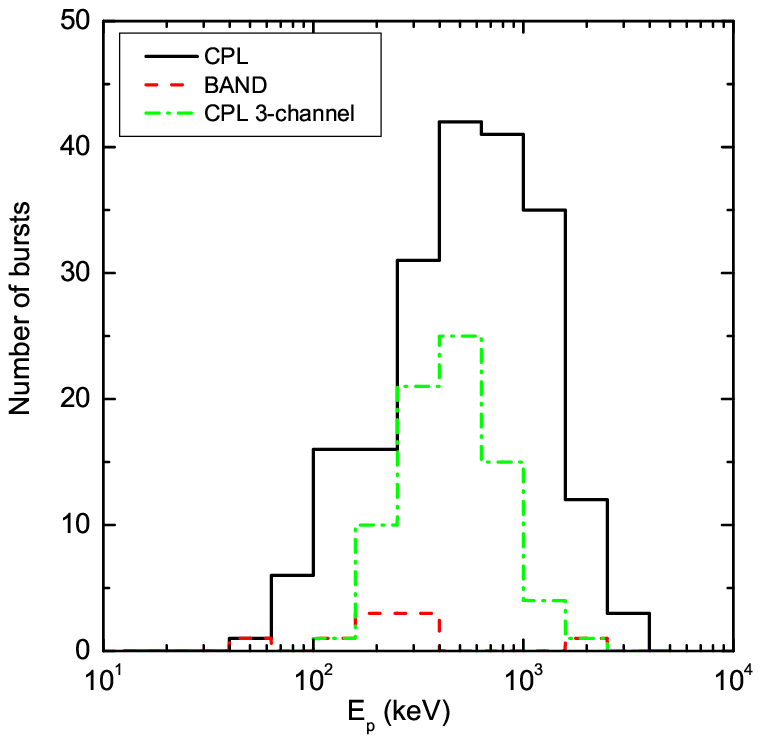}}
	\end{center}
    \caption{Distributions of $\alpha$~\subref{fig:alpha} and $E_\rmn{p}$~\subref{fig:Ep}
        obtained from time-integrated spectral fits with different models,
        shown in colors. \label{fig:par_dist} }
\end{figure}

In Figure~\ref{fig:par_dist}, we show the distributions for $E_\rmn{p}$ and $\alpha$.
The low-energy indices $\alpha$ of the best-fit models for the multichannel spectra
are distributed around a value of~$-0.5$.
About 66\% of the low-energy indices are $\alpha > -2/3$, violating the synchrotron ``line-of-death''~\citep{Preece_1998ApJL},
while only 1\% of the indices (three photon indices of the PL model) are $\alpha< -3/2$,
violating the synchrotron cooling limit.
For the four spectra that are best described with the PL model the photon indices are at the soft
end of the low-energy index distribution. The high-energy indices are distributed around a slope $\beta=-2.3$.
The $E_\rmn{p}$ distribution for the CPL model peaks around 500~keV and covers
just over two orders of magnitude.
We studied the difference in the value of $E_\rmn{p}$ between the BAND and CPL fits in the GOOD sample.
We found that for each spectrum the $E_\rmn{p}$ in the CPL and BAND models in the GOOD sample
are consistent within 90\% CL.

Three bright GRBs are found to have $P<0.001$ for the fits of the time-integrated spectra.
We have explored these GRBs in more detail.
In the case of GRB20060306\_T55358, with $P \approx 10^{-4}$, the strong hard-to-soft
evolution of the emission results in the poor BAND and CPL fits to the time-integrated spectrum.
However, we found no strong deviation from the BAND model ($P>0.05$) for the individual,
time resolved spectra of this bright GRB.
The time-integrated spectra of two other bursts, GRB19960908\_T25028 and GRB20031214\_T36655
turn out to be well described by a sum of CPL and PL functions.
In addition, we found apparent systematics in the fit residuals for GRB19980205\_T19785 ($P=0.08$)
whose spectrum is also well described by the CPL+PL combination.
The parameters of the CPL+PL fits to time-integrated spectra of these GRBs are given in Table~\ref{tab:extra_comp}.
For all three GRBs the PL component, which is also detected in most of
the time resolved spectra of these bursts, is rather soft ($\alpha \sim -2$)
and dominates the emission below $\sim 50$--100~keV.
The hard CPL component is described by $E_\rmn{p}\sim(1.5\textrm{--}2)$~MeV and
considerably flatter photon index ($\alpha > -1$).
All the above-mentioned GRBs are in the top 10\% of the most intense ones
in terms of their energy fluence, with GRB20031214\_T36655 and GRB20060306\_T55358
being the first and the second most intense bursts in our sample, respectively.

\subsection{Fluences and peak fluxes}
The values of $S$ and $F_\rmn{peak}$ were derived
using the energy flux of the best-fit spectral model in the 10~keV--10~MeV band.
Since the spectrum accumulation interval typically differs from the $T_{100}$ interval
a correction which accounts for the emission outside the time-integrated spectrum
was introduced when calculating $S$.
For short GRBs with EE, the energy fluences of the initial peak and EE were 
estimated separately (see also Section~\ref{sec:EE}).
$F_\rmn{peak}$ was calculated on the 16~ms scale using the best-fit spectral
model for the spectrum near the peak count rate.
To obtain $F_\rmn{peak}$, the model energy flux was multiplied
by the ratio of the 16~ms peak count rate to the average count rate in the spectral accumulation interval.
Typically, the corrections were made using counts in the G2+G3 light curve;
the G1+G2, G2 only, and G1+G2+G3 combinations were also considered depending on
the emission hardness and intensity.

\begin{figure}
	\begin{center}
		\subfigure[]{\label{fig:fl}\includegraphics[height=0.45\textwidth]{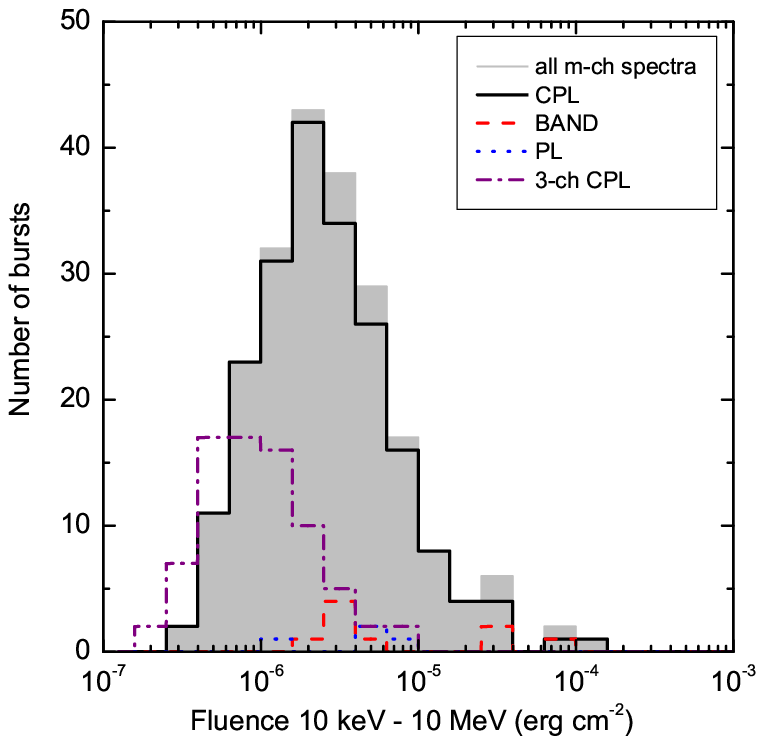}}
		\subfigure[]{\label{fig:pf}\includegraphics[height=0.45\textwidth]{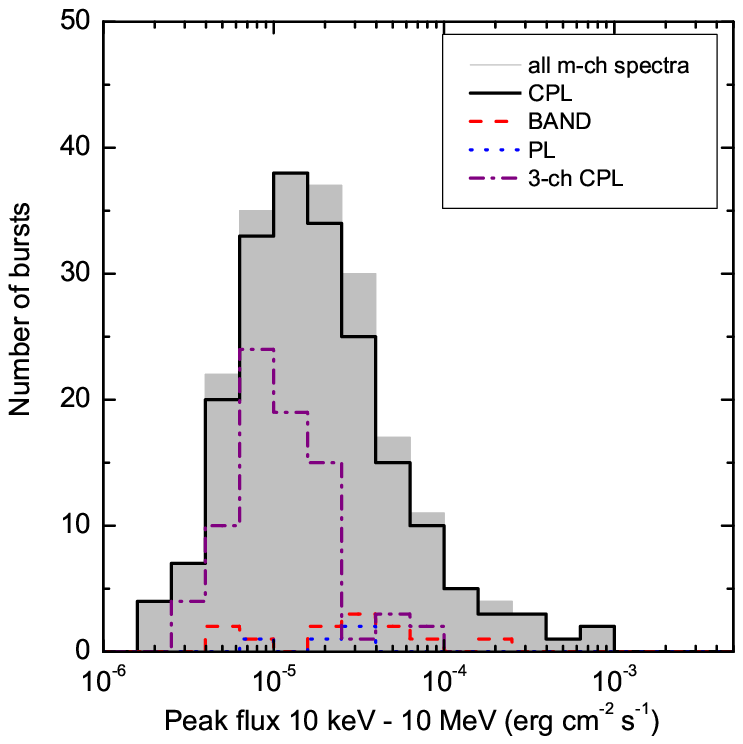}}
	\end{center}
\caption{Distributions of the total energy fluence~\subref{fig:fl} and
        the peak energy flux~\subref{fig:pf}.
        The gray filled histogram in each panel shows the total distribution for 214 multichannel spectra
        and the constituents are shown in color.
        The dash-dotted histograms show distributions for 79 three-channel spectra.\label{fig:fl_pf_dist} }
\end{figure}

Table~\ref{tab:fl_pf} contains $S$ and $F_\rmn{peak}$ for the 293 bursts.
The first column gives the burst designation (see Table~\ref{tab:info}).
The three subsequent columns give $S$;
the start time of the 16-ms time interval, when the peak count rate in
the G2+G3 band is reached; and $F_\rmn{peak}$. The distributions of
$S$ and $F_\rmn{peak}$ are shown in Figure~\ref{fig:fl_pf_dist}.
The ranges of $S$ and $F_\rmn{peak}$ are $(0.2\textrm{--}140)\times 10^{-6}$~erg~cm$^{-2}$
and $(0.2\textrm{--}85)\times 10^{-5}$~erg~cm$^{-2}$~s$^{-1}$, respectively.

We note that for the handful of very intense, highly variable GRBs
(i.e. GRB19970704\_T04097, GRB20031214\_T36655, GRB20051103\_T33943, GRB20060306\_T55358,
GRB20070201\_T55390, and GRB20070222\_T27115)
the values of $F_\rmn{peak}$ (and to a lesser extent $S$) can be underestimated in our analysis
by a factor of $\sim1.5$--2, because the live time in a spectrum
is estimated under the assumption of a constant count rate during the accumulation interval.

\subsection{Short GRBs with extended emission}\label{sec:EE}
The extended emission (EE) component which follows the initial short
pulse (IP) has been observed in a number of short GRBs by various experiments:
\textit{CGRO}-BATSE~\citep{Burenin_2000AstL, Norris_and_Bonnel_2006ApJ, Bostanci_2013MNRAS},
KW~\citep{Mazets_2002astroph, Frederiks_2004_EE},
\textit{INTEGRAL}-SPI-ACS~\citep{Minaev_2010AstL},
\textit{Swift}-BAT~\citep{Norris_2011ApJ, Sakamoto_2011ApJS},
and \textit{Fermi}-GBM~\citep{Kaneko_2015MNRAS}.
We searched for candidates for short GRBs with EE in the full sample of \ntot \kw GRBs
detected between 1994 and 2010. We defined the following search criteria:
the burst initial pulse should meet our criteria for a short GRB, i.e. have $T_{50}<0.6$~s;
and the remaining part of burst (EE) should not exhibit peaks with prominent spectral evolution.
Applying these criteria to the full KW sample, we found 31 candidates for short GRBs with EE.
Although the bright IP of GRB~070207 \citep{Golenetskii_2007GCN6089} satisfies our criteria
of a short GRB with $E_\rmn{p} \sim 300$~keV, the very intense and spectrally-hard ($E_\rmn{p} \sim 1.5$~MeV) 
behavior of the subsequent emission, which only formally can be considered as EE,
suggests that this event is a long-duration, hard-spectrum burst with a short GRB-like precursor,
very similar in morphology to two other KW bursts, GRB~000115 and GRB~001020.

Only for 21 of the remaining 30 events was the EE bright enough to allow spectral analysis.
The initial pulses of these events are classified in Table~\ref{tab:durations} as Iee or Iee/II.
Table~\ref{tab:EE} presents the parameters of the EE. The ten columns contain the following
information: (1) the burst designation (see Table~\ref{tab:info});
columns (2) and (3) contain the EE start time (relative to $T_0$) and duration,
determined at the $5\sigma$ confidence level in the G2 or G2+G1 bands;
columns (4) and (5) contain the spectrum start time $T_\rmn{start}$ (relative to $T_0$)
and its accumulation time~$\Delta T$;
(6) best-fit models for each spectrum;
(7) and (8) contain $\alpha$ and $E_\rmn{p}$;
(9) the EE energy fluence in the 10~keV--10~MeV band;
(10) $\chi^2/\rmn{dof}$ along with the null hypothesis probability~$P$.

In 15 cases EE is best fitted with PL and in six cases with the more complex CPL model. 
The PL indices range from $-2.6$ to~$-1.4$ with a median of~$-1.6$, 
and the CPL photon indices range from~$-1.4$ to~$-0.3$ with a median of~$-1.2$. 
The $E_\rmn{p}$ values range from $\approx 160$~keV to $\approx 2.2$~MeV
with a median of $\sim 300$~keV and a geometric mean of 370~keV.
For the 21 bursts, the fluence ratio, EE to initial pulse, ranges from 
0.06 to 15 with a median of 3.3. 
Among six KW bursts whose EE can be well described with the CPL model, four have the $E_\rmn{p}$ of 
the EE lower than that of the IP. Two bursts, GRB19950526\_T16613 and GRB20090720\_T61379,
display EE harder than IP, with the latter having extremely hard EE ($E_\rmn{p} = 2.2(-1.0,+2.4)$~MeV).

\section{SUMMARY AND DISCUSSION}\label{sec:SUMMARY}

We have presented the results of the systematic spectral analysis of 293 short \kw GRBs,
which is $\sim 15$\% of all \kws GRBs detected during the first fifteen years of operation.
Among them, $\sim 70$\% are classified as Type~I bursts, $\sim 8$\% as Type~II ,
and $\sim 12$\% have an uncertain type (I or~II).
The fraction of \kws short GRBs that display extended emission is $\sim 10$\%.

In total we analyzed 253 multichannel spectra: 214 time-integrated spectra,
18 spectra near the peak count rate, and 21 spectra of the extended emission.
We also analyzed 79 three-channel spectra.
Table~\ref{tab:pardist} contains the median values and 90\% confidence intervals (CIs) for
spectral parameter and energetics distributions resulting from our analysis.
The first column gives the model name. The second gives the spectral data type:
multichannel or three-channel. The subsequent ten columns contain median
parameter values and 90\% CI for $\alpha$, $\beta$, $E_\rmn{p}$, $S$, and $F_\rmn{peak}$.
The highest $E_\rmn{p}$ found for KW short GRBs are $\sim 3$~MeV:
$E_\rmn{p} = 3.55(-0.71,+0.85)$~MeV was observed in GRB20090510\_T01381
(GRB 090510; \citealp{Ackermann_2010ApJ_716_1178A});
just slightly softer are GRB19970704\_T04097 and GRB20080611\_T04742, both with $E_\rmn{p} \approx 3.3$~MeV.
Almost all GRBs with $E_\rmn{p} \lesssim 200$~keV are classified as Type~II or Type~I/II
and probably represent a population different from that of the harder GRBs (see the discussion below).

Our results support the previous findings that the spectra of the majority of short GRBs are well
described by the CPL function with hard $\alpha \sim -0.5$ and $E_\rmn{p}$ in the range of 100~keV--2~MeV.
We found that the Band function is the best-fit model only for $\sim 4$\% of \kws short GRBs.
Among the 5\% highest-$S$ GRBs 20\% are best-fitted with BAND; the remaining 80\% of the bursts
require a high-energy index $\beta \lesssim -2.5$ and in most cases are not constrained from below.
This suggests that the absence of high-energy PL behavior observed in a large fraction
of the bright short GRB spectra is likely intrinsic to the bursts rather than due to poor count statistics.

The scope of this catalog does not involve a study of the short GRB spectra with more complex models.
Nevertheless, we found, that among the 214 bursts with multichannel spectra,
three GRBs require an additional PL component with photon index of $\sim -2$.
These bursts belong to the brightest 10\% of the sample.
The ratio of the PL to CPL component energy flux ranges from 0.03 in GRB20031214\_T366655
to 0.4 in GRB19980205\_T19785.
These PL components might be similar to
that found for GRB~081024B~\citep{Abdo_2010ApJ_712_558A} and
GRB~090510~\citep{Ackermann_2010ApJ_716_1178A} using \textit{Fermi}~GBM and~LAT data.
GRB~081024B was not detected by \kws in the triggered mode while GRB~090510 (GRB20090510\_T01381)
is present in our sample. For the latter burst, the additional PL component is
not needed to describe the \kws time-integrated spectrum, and
the estimated upper limit to the energy flux of the PL component with
a photon index of $-1.7$ is $\sim 1 \times 10^{-6}$~erg~cm$^{-2}$~s$^{-1}$ at 90\%~CL.
The corresponding energy flux ratio, PL to BAND, is less than $\sim 0.02$ at 90\%~CL. 
A detailed study of the \kws bursts with an additional spectral component
will be published in a separate paper.

\subsection{Comparison of KW with BATSE and GBM short GRBs}
We compared the results of our spectral analysis to those reported for
other instruments.
The largest broadband GRB samples available to date are those reported by 
\textit{CGRO}-BATSE\footnote{http://heasarc.gsfc.nasa.gov/W3Browse/cgro/bat5bgrbsp.html}
(20~keV--2~MeV; \citealp{Goldstein_2013ApJS}) and 
\textit{Fermi}-GBM\footnote{http://heasarc.gsfc.nasa.gov/W3Browse/fermi/fermigbrst.html}
(8~keV--40~MeV; \citealp{Gruber_2014ApJS}). 
The present analysis contains about a factor of two more short GRBs than the GBM study,
over a slightly narrower energy range, and a factor of $\sim 1.5$ less than the BATSE sample,
but over a broader energy range.

From the BATSE 5B catalog we selected 427 bursts with $T_{90}<2$~s and
with the time-integrated spectrum accumulation interval being shorter than 10~s.
Based on a $\Delta \chi2>6$ criterion, the best-fit model statistics for these GRBs is:
11~--- Band, 225~--- CPL, and 191~--- PL.
From the GBM second catalog we selected 146 GRBs with $T_{90}<2$~s.
The best-fit models for the GBM bursts, as given in the catalog, are:
3~--- Band, 67~--- CPL, and 76~--- PL;
the bursts best described with smoothly broken power law were excluded from the comparison.

The ratio of Band to CPL best-fit models is small ($\lesssim 5$\%) for all samples.
We tested whether distributions of $\alpha$ and $E_\rmn{p}$ of the CPL model are
consistent between the instruments. The two-sided p-values
of the two-sample Kolmogorov-Smirnov test ($P_\rmn{KS}$) for KW and GBM $\alpha$
and $E_\rmn{p}$ distributions are 10\% and 25\%, respectively, while for KW and BATSE,
and GBM and BATSE $P_\rmn{KS}<1$\%. The BATSE sample has median $\alpha=-0.33$
while medians for KW and GBM are $\alpha=-0.49$ and $\alpha=-0.50$, respectively.
The median of the $E_\rmn{p}$ distribution is $\sim 400$~keV for BATSE and $\sim 550$~keV for both KW and GBM.
Thus, the KW results are consistent with GBM and to a slightly lesser extent with BATSE.

The fractions of best-fit PL models in each sample are: 
2\% (5\% using  $\Delta \chi^2 > 6$ criterion)~--- KW, 52\%~--- GBM, 55\%~--- BATSE.
We have investigated the origin of a high fraction of PLs in the BATSE and GBM samples.
To make the comparison more robust, we selected the bursts with $S>5.5\times 10^{-7}$~erg~cm$^{-2}$,
which is approximately the lowest $S$ measured for KW short GRBs with multichannel spectra.
The resulted subsamples of 138 (BATSE) and 49 (GBM) GRBs contain 29 (21\%) and 3 (6\%) PLs, respectively.
Thus, in the common fluence range, the fractions of best-fit PLs for KW and GBM are consistent.
Among GOOD CPL models for the 29 BATSE bursts 17 have $1\sigma$ upper limits of $E_\rmn{p}$
not constrained to the upper BATSE spectral band boundary (2~MeV). 
The remaining 12 bursts represent 9\% of the subsample.
Thus, the main source of the relatively high PL fractions in the BATSE and GBM short GRBs samples 
is a large amount of weak bursts for which a more complex model cannot be preferred
due to low count statistics. Also, for the BATSE bursts the additional bias toward the PLs
comes from the relatively narrow energy band.

\subsection{Extended emission}
We found that 30 bursts from the sample of \ntot \kws GRBs detected from 1994 to 2010
can be classified as short GRBs with EE based on the short duration of
an initial pulse and the presence of subsequent emission exhibiting
no prominent spectral evolution. Of them 21 GRBs have intense enough EE to
perform spectral analysis. 
For six KW bursts the EE spectrum requires a ``curved'' (CPL) model with 
rather high $E_\rmn{p} \sim 160$~keV--2.2~MeV. The IPs of two of them are classified as Iee/II 
and they are probably long GRBs with a short initial pulse. 
The four remaining events, however, are ``canonical'' short/hard GRBs with
EE in terms of the light curve shape. 
Similar EE spectral behavior was reported earlier for two out of 19 BATSE 
GRBs~\citep{Bostanci_2013MNRAS} and for four out of 14 GBM bursts~\citep{Kaneko_2015MNRAS};
our results provide additional evidence of rather hard EE being observed in some short GRBs.

In total, our sample contains two short GRBs with EE detected by BATSE 
and three EE bursts detected by GBM. The comparison of the fits shows that the EE spectral parameters 
for these bursts, including those for one common GRB with hard EE (GRB20090831\_T27393; $E_\rmn{p} \approx 215$~keV), 
are consistent within errors between KW and the other instruments.
The KW GRB20090720\_T61379 showing extremely hard EE ($E_\rmn{p} \approx 2.2$~MeV) 
was also detected by GBM. Although this burst had not been included by~\citet{Kaneko_2015MNRAS} in the EE sample, 
the GBM time-integrated spectral parameters~\citep{Gruber_2014ApJS} are consistent 
with the KW fits we made for the same time interval.

The bright, nearby GRB~060614, which can be regarded as a short GRB with EE~\citep{Gehrels_2006Nature},
was detected by KW ($T_0$(KW)=45831.590~s;~\citealp{Golenetskii_GCN5264}),
but was not included in our short GRBs sample because of the long duration
of the initial peak $T_{50}=2.7 \pm 0.3$~s.

\subsection{Giant flare candidates}
The enormous initial pulse of a soft gamma-repeater giant flare (GF) can mimic
a classical short GRB even when observed from a nearby galaxy.
An upper limit on the fraction of such events among observed short GRBs was
estimated in several studies to be $\sim 1$--15\%, see~\citet{Hurley_2011ASR} for a review.
\citet{Svinkin_2015MNRAS} performed a search for GFs in the \kws short GRB sample
using the burst localizations from~\citet{Palshin_2013ApJS}.
Only two earlier reported candidates were found, GRB~051103 (GRB20051103\_T33943)
in the M81/M82 group of galaxies~\citep{Frederiks_2007AstLett} and
GRB~070201 (GRB20070201\_T55390) in the Andromeda galaxy~\citep{Mazets_2008ApJ}.
Both GRB~051103 and GRB~070201 are in the 10\% of the most intense bursts
in terms of total energy fluence, while GRB~051103 is the brightest in terms of the peak energy flux.
The spectral parameters of the bursts are typical for our sample.
A count excess observed up to 90~s after trigger for GRB~070201
was suggested by~\citet{Mazets_2008ApJ} to be the tail of the possible GF.
The significance of the excess is $4.3\sigma$ and it does not meet our $5\sigma$ EE criterion.

\subsection{Heterogeneity of short GRBs}
\begin{figure}
	\begin{center}
		\subfigure[]{\label{fig:EpTypeII}\includegraphics[height=0.45\textwidth]{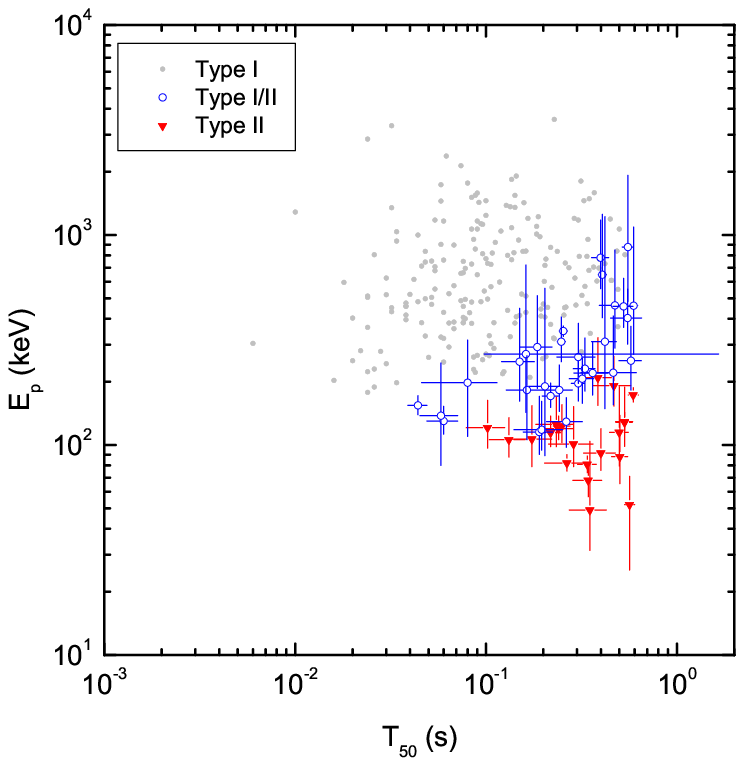}}
		\subfigure[]{\label{fig:EpEEip}\includegraphics[height=0.45\textwidth]{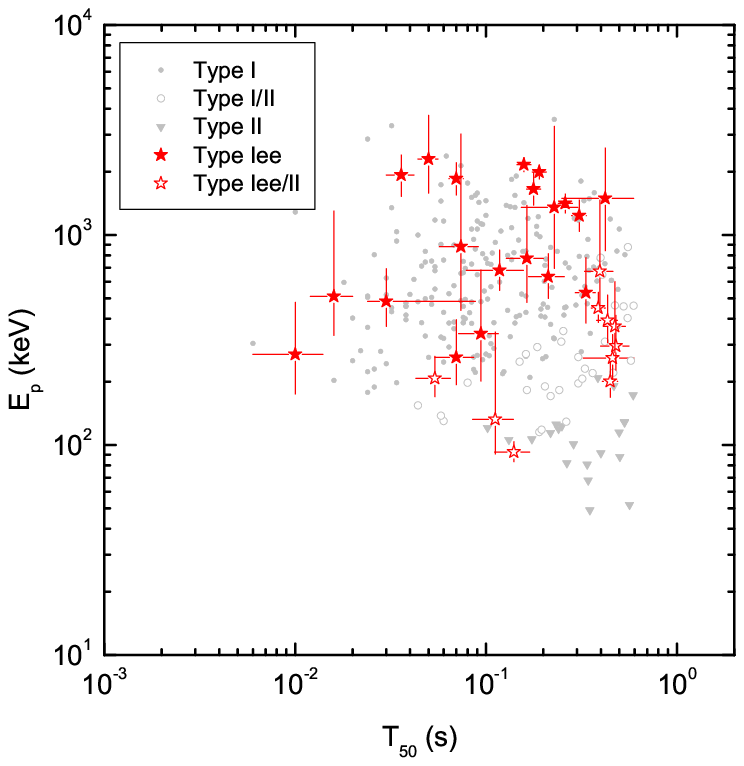}}
	\end{center}
\caption{Best-fit model $E_\rmn{p}$ as a function of $T_{50}$ (only CPL fits are shown).
        Panel~\subref{fig:EpTypeII} shows Type~I bursts (gray dots), Type~II bursts (red triangles),
        and bursts of uncertain type, I or~II (blue circles).
        Panel~\subref{fig:EpEEip} shows the Type~I bursts with EE (filled red stars),
        and bursts of uncertain type, Iee or~II (empty stars).
        For Type~I GRBs error bars are not shown.
        \label{fig:EpT50}}
\end{figure}

Figure~\ref{fig:EpT50} shows $E_\rmn{p}$ of the CPL best-fit model
as a function of the burst duration $T_{50}$.
The Type~I GRBs tend to be harder ($E_\rmn{p}\gtrsim 200$~keV) and shorter than Type~II bursts,
which is consistent with the classification obtained using the hardness-duration distribution.
Among four bursts best-fitted with the PL model, two are types~I and~II,
and two have uncertain classifications (I/II).
The apparent lack of \kws GRBs with $E_\rmn{p}\lesssim 100$~keV and $T_{50}\lesssim 0.3$~s
is probably due to selection effects.
The duration distribution of the initial pulses of short GRBs with EE (Iee) is consistent
with that of the Type~I bursts; this is supported by $P_\rmn{KS} \sim 0.5$.
We found that the $E_\rmn{p}$ of the initial pulses of Iee bursts are, on average,
harder than the $E_\rmn{p}$ of the Type~I bursts by a factor of $\sim 1.5$, and $P_\rmn{KS}$
for the two $E_\rmn{p}$ distributions is $\sim 0.01$.
Finally, we tested whether the $S$ and $F_\rmn{peak}$ distributions for the initial peaks of the Iee bursts
differ from those of Type~I GRBs. In both cases we obtained $P_\rmn{KS} \sim 0.01$
which disfavors the hypothesis that both Iee and Type~I GRBs are drawn from the same population,
with short GRBs with EE being, on average, more intense.

\begin{figure}
	\begin{center}
		\begin{minipage}[t]{1\textwidth}
		\subfigure[]{\label{fig:EpFluence}\includegraphics[width=0.5\textwidth]{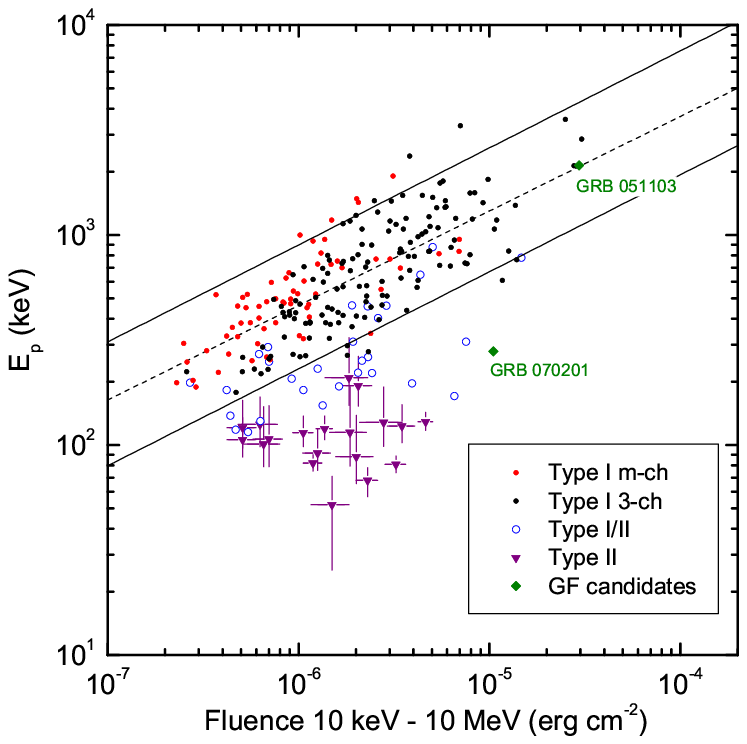}}
		\subfigure[]{\label{fig:EpFluence_EE}\includegraphics[width=0.5\textwidth]{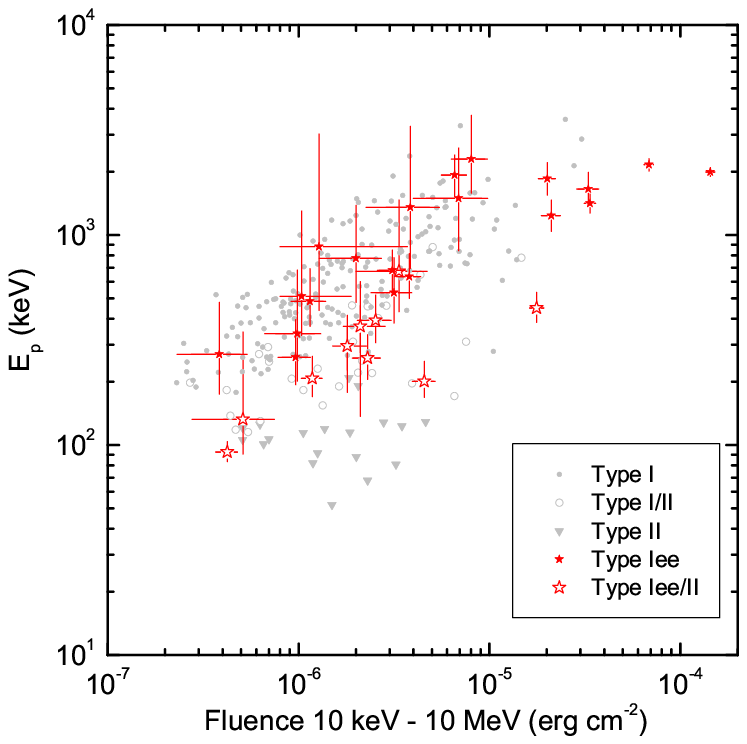}}
		\end{minipage}
		\begin{minipage}[t]{1\textwidth}
		\subfigure[]{\label{fig:EpPF}\includegraphics[width=0.5\textwidth]{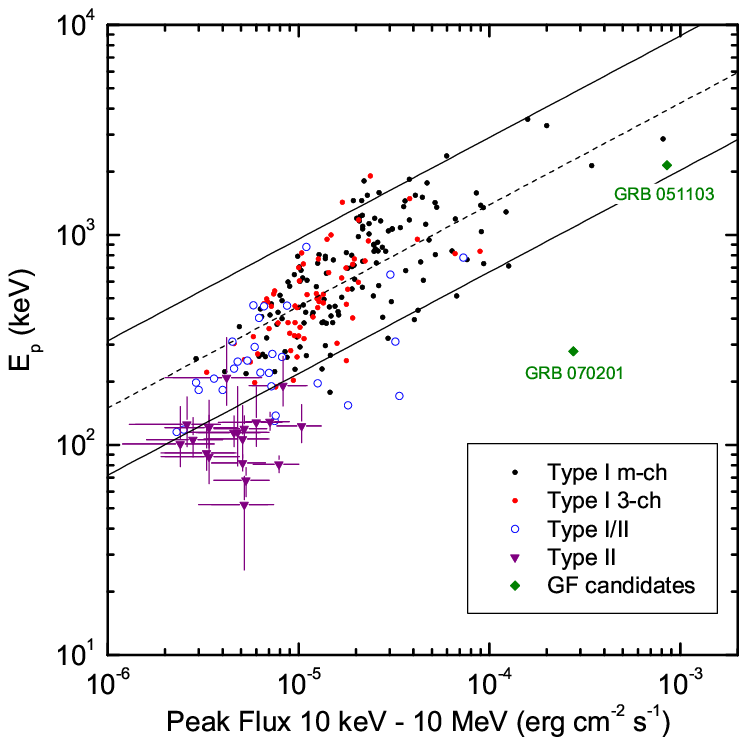}}
		\subfigure[]{\label{fig:EpPF_EE}\includegraphics[width=0.5\textwidth]{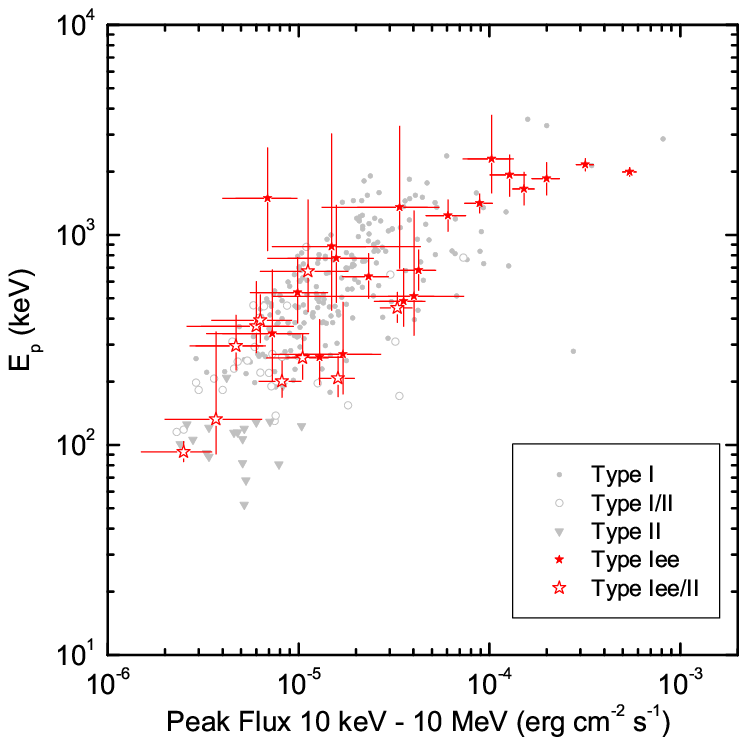}}
		\end{minipage}
	\end{center}
\caption{\scriptsize
$E_\rmn{p}$ as a function of the total energy fluence and the peak energy flux for the best CPL fits.
Panel~\subref{fig:EpFluence} shows $E_\rmn{p}$ vs. the total energy fluence distribution for
the Type~I bursts with multichannel spectra (black circles);
the Type~I bursts with three-channel spectra (red circles);
the bursts with uncertain type (empty circles), for both types of spectra;
and the Type~II bursts (triangles).
Panel~\subref{fig:EpFluence_EE} shows bursts of types Iee (filled  stars) and~Iee/II (empty stars);
the remaining bursts from the sample are shown in gray.
Panels~\subref{fig:EpPF} and~\subref{fig:EpPF_EE} show $E_\rmn{p}$ vs. the peak energy flux
distribution for the same GRB groups. For the GRBs of type~I and~I/II error bars are not shown.
The extragalactic SGR giant flare candidates are shown with diamonds.
The dashed lines denote the best powerlaw fits for the $E_\rmn{p}$--$S$ (with an index of $0.46\pm0.16$)
and $E_\rmn{p}$--$F_\rmn{peak}$ (with an index of $0.48\pm0.18$) relations
of the Type~I GRBs. The solid lines denote the 90\% prediction bands.
\label{fig:EpvsFPandFL}}
\end{figure}

Figure~\ref{fig:EpvsFPandFL} shows $E_\rmn{p}$ as a function of $S$ and $F_\rmn{peak}$.
The 79 faint bursts, for which only three-channel spectra were available for the analysis,
show spectral parameter and $F_\rmn{peak}$ distributions similar to those
of more fluent GRBs from the sample; also, these bursts smoothly extend
the short-hard (Type~I) GRB distribution to the low-$S$ region in the $E_\rmn{p}$--$S$ plane.
The candidate for a GF in the Andromeda galaxy (GRB~070201) is an apparent outlier
in the $E_\rmn{p}$--$F_\rmn{peak}$ distribution, supporting the non-GRB nature of this event.
Type~I and Type~II bursts occupy
virtually non-overlapping regions in the $E_\rmn{p}$--$S$ plane.
The Type~I GRBs form an elongated distribution that generally follows an $E_\rmn{p} \propto S^{1/2}$ relation.
The Type~II bursts from our sample do not share this correlation; they form a small,
soft-spectrum population which represents a tiny fragment of the long-soft \kws GRB distribution.
To a lesser extent, the same is true of the Type~I and Type~II population
behavior in the $E_\rmn{p}$--$F_\rmn{peak}$ plane.
Since only nine bursts from our sample have known redshift
($z\sim 0.1$--1.0, determined either spectroscopically or photometrically),
the rest-frame properties of the bursts are not discussed in this work.
The detailed analysis of all \kws GRBs with known redshifts will be presented
in a separate paper (Tsvetkova~et~al.,~in prep.).
Although the instrumental biases affect the burst sample properties,
the correlations in the observer frame may still be the consequences of
the rest-frame $E_\rmn{p,rest}$--$E_\rmn{iso}$~\citep{Amati_2002AandA}
and $E_\rmn{p,rest}$--$L_\rmn{iso}$~\citep{Yonetoku_2004ApJ} correlations
(see, e.g.,~\citealp{Nava_2008MNRAS} and references therein).
Thus, the properties of the observer-frame hardness-intensity distribution
obtained for Type I and Type II bursts from the KW short GRB sample favor the hypothesis that
short/hard GRBs follow their own form of the ``Amati'' $E_\rmn{p,rest}$--$E_\rmn{iso}$ relation
(see, e.g.,~\citealp{Nava_2011MNRAS} and references therein).

\begin{figure}
	\begin{center}
		\subfigure[]{\label{fig:logNlogS}\includegraphics[height=0.45\textwidth]{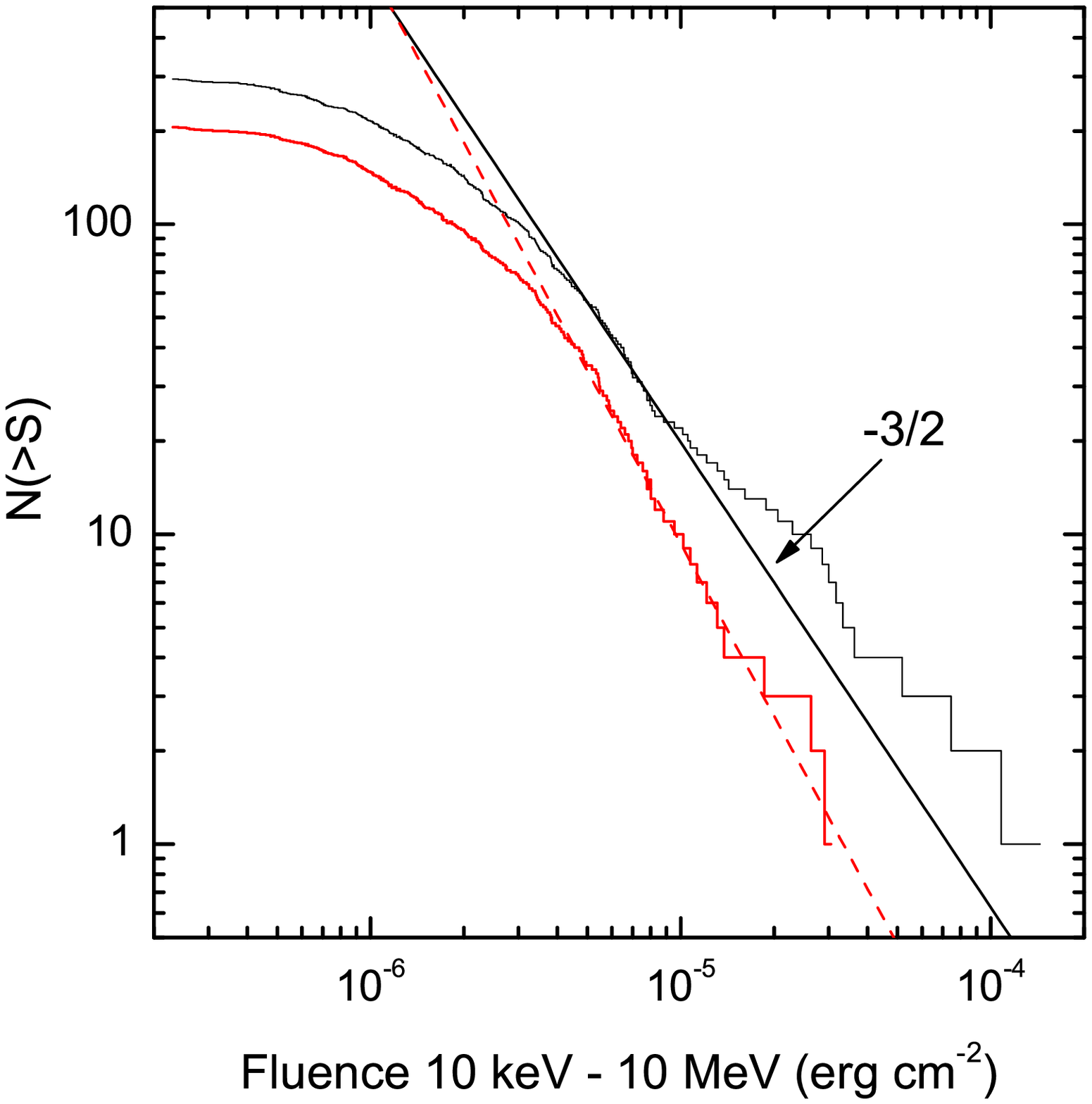}}
		\subfigure[]{\label{fig:logNlogPF}\includegraphics[height=0.45\textwidth]{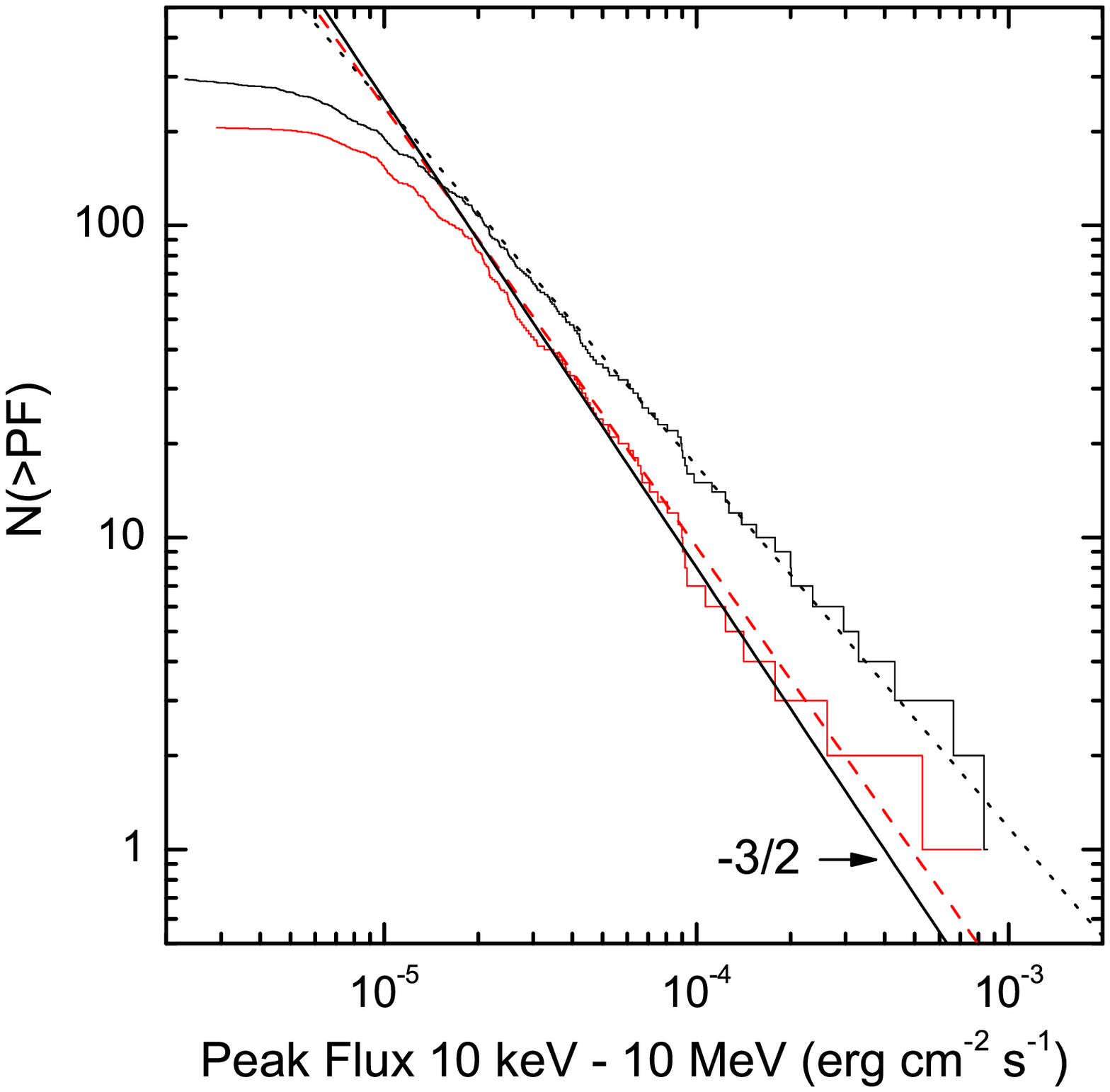}}
	\end{center}
\caption{Cumulative distributions of the total energy fluence (left panel)
        and the peak energy flux (right panel).
        In both panels upper (black) and lower (red) histograms represent the distributions
        built for the whole short GRB sample and for the Type~I GRB sub-sample, respectively.
        The dashed lines and dotted lines show the best power-law approximations to the corresponding distributions
        (see Section~\ref{sec:SUMMARY} for the approximation ranges and the indices).
        The solid lines show a slope of $-3/2$ expected if GRBs were homogeneously distributed
        in an Euclidean space throughout the sampled volume.
        \label{fig:logNlogS_PF} }
\end{figure}

Figure~\ref{fig:logNlogS_PF} shows $\log N$--$\log S$ and $\log N$--$\log F_\rmn{peak}$ 
distributions for 293 \kws short GRBs
along with a homogeneous space distribution with index $-3/2$.
The $\log N$--$\log S$ distribution tends to follow this slope only in a limited
range of fluences, $(\sim4\textrm{--}10)\times 10^{-6}$~erg~cm$^{-2}$.
While the deficit of the faint bursts can be explained by instrumental bias,
the visible excess of intense bursts is, to a significant extent, due to events 
not representing the ``classical'' short/hard GRB population.
Among the 12 most energetic GRBs in our sample, with $S\gtrsim 2\times 10^{-5}$~erg~cm$^{-2}$,
only four are of Type~I; the others are Type~I/II, Type~II, or bursts with EE (Iee).
After all non-Type~I GRBs are excluded from the consideration, the $\log N$--$\log S$ distribution
shows a good agreement with a steep slope of $-1.85 \pm 0.30$ above $S\sim 5\times 10^{-6}$~erg~cm$^{-2}$.
The $\log N$--$\log F_\rmn{peak}$ distribution of the KW short GRBs is also more shallow than
the $-3/2$ slope;  
we estimate the power-law index
of the integral distribution to be $-1.16 \pm 0.12$ for $F_\rmn{peak}$ in the
$(0.2\textrm{--}9.4)\times 10^{-4}$~erg~cm$^{-2}$~s$^{-1}$ range.
In the same range, the $\log N$--$\log F_\rmn{peak}$ distribution of Type~I bursts
demonstrates a steeper slope of $-1.42 \pm 0.16$, in agreement
with the homogenous distribution.

Plots of the \kws short GRBs time histories and spectral fits can be found
at the Ioffe Web site\footnote{http://www.ioffe.ru/LEA/shortGRBs/Catalog2/}.
We note that \kws continues to operate well, and has detected $\sim 380$ short bursts up to December 2015.
The results of the analyses of the short GRBs detected by \kw after 2010 will
be presented on-line at the same URL.


\acknowledgments
We thank the reviewer comments which significantly contributed to improving
the quality of the publication.
R.L.A. and S.V.G. gratefully acknowledge support from RFBR grants 15-02-00532 and 13-02-12017-ofi-m.
This research made use of Astropy\footnote{http://www.astropy.org}, 
a community-developed core Python package for Astronomy~\citep{Robitaille_2013AandA}.

{\it Facilities:} \facility{\textit{Wind} (Konus)}.


\bibliography{biblio}

\clearpage

\clearpage
\begin{deluxetable}{cccccc}
\tabletypesize{\scriptsize}
\tablecaption{\kw short GRB observation details\label{tab:info}}
\tablewidth{0pt}
\tablehead{
\colhead{Designation} & 
\colhead{\kw} &
\colhead{Name\tablenotemark{a}} &
\colhead{Detector} &
\colhead{Incident angle} & 
\colhead{Comment\tablenotemark{b}}\\
\colhead{} & 
\colhead{Trigger Time (UT)} & 
\colhead{} & 
\colhead{} & 
\colhead{(\degr)} & 
\colhead{} 
}
\startdata   
 GRB19950210\_T08424 & 02:20:24.147  &    \nodata & S1 &  55(-0,+0)  & 2 \\
 GRB19950211\_T08697 & 02:24:57.748  &    \nodata & S2 &  47(-0,+0)   & 2 \\
 GRB19950414\_T40882 & 11:21:22.798  &    \nodata & S1 &  57(-57,+30) & 5 \\
 GRB19950503\_T66971 & 18:36:11.838  &    \nodata & S1 &  73(-0,+0)   & 2 \\
 GRB19950520\_T83271 & 23:07:51.403  &    \nodata & S1 &  46(-46,+30) & 5 \\
\enddata
\tablenotetext{a}{as provided in the GCN circulars, if available}
\tablenotetext{b}{                                                                
    1~--- detected by imaging insruments (incident angle error is not given);
    2~--- burst localalized to a box or segment, localization center is used;
    3~--- burst localalized to a box or segment, ecliptic latitude estimate is used;
    4~--- burst localalized to a single annulus, a position on the annulus center line 
         which is the most consistent with the ecliptic latitude estimate is used;
    5~--- observed by \kw only, ecliptic latitude estimate is used.}
\tablecomments{(This table is available in its entirety in a machine-readable form in the online journal. 
A portion is shown here for guidance regarding its form and content.)}
\end{deluxetable} 
\begin{deluxetable}{ccccccccc}
\tabletypesize{\scriptsize}
\tablecaption{Durations, spectral lags, and classification\label{tab:durations}}
\tablewidth{0pt}
\tablehead{
\colhead{Designation} & 
\colhead{$t_0$\tablenotemark{a}} &
\colhead{$T_{100}$} &
\colhead{$T_{50}$} &
\colhead{$T_{90}$} &
\colhead{Type} &
\colhead{$\tau_\rmn{lag32}$} &
\colhead{$\tau_\rmn{lag31}$} &
\colhead{$\tau_\rmn{lag21}$} \\
\colhead{}    & 
\colhead{(s)} & 
\colhead{(s)} & 
\colhead{(s)} & 
\colhead{(s)} & 
\colhead{}    & 
\colhead{(ms)} & 
\colhead{(ms)} & 
\colhead{(ms)}   
}
\startdata   
 GRB19950210\_T08424 & $-0.034$ & 0.176 & $0.060 \pm 0.011$ & $0.138 \pm 0.024$ &   I/II & \nodata     & \nodata     & \nodata     \\
 GRB19950211\_T08697 & $-0.046$ & 0.214 & $0.030 \pm 0.004$ & $0.104 \pm 0.030$ &      I & $12 \pm  2$ & $22 \pm 10$ & $18 \pm 8$  \\
 GRB19950414\_T40882 & $-0.130$ & 0.164 & $0.048 \pm 0.020$ & $0.150 \pm 0.014$ &      I & $ 6 \pm 10$ & \nodata     & \nodata     \\
 GRB19950503\_T66971 & $-0.206$ & 0.402 & $0.050 \pm 0.006$ & $0.282 \pm 0.049$ &    Iee & $-1 \pm  3$ & \nodata     & $-1 \pm 14$ \\
 GRB19950520\_T83271 & $-0.162$ & 1.218 & $0.210 \pm 0.073$ & $1.100 \pm 0.219$ &      I & $1  \pm 13$ & \nodata     & \nodata     \\
 GRB19950526\_T16613 & $-0.510$ & 1.934 & $0.480 \pm 0.081$ & $1.464 \pm 0.377$ & Iee/II & \nodata     & \nodata     & \nodata     \\
\enddata
\tablenotetext{a}{Relative to the trigger time}
\end{deluxetable}

\begin{landscape}
{\renewcommand\tabcolsep{3pt}
\begin{deluxetable}{cccrrrcccc}
\tabletypesize{\scriptsize}
\tablecaption{Spectral parameters (multichannel spectra)\label{tab:spec_par}}
\tablewidth{0pt}
\tablehead{
\colhead{Designation} & 
\colhead{Spec.} &
\colhead{$T_\rmn{start}$\tablenotemark{a}} &
\colhead{$\Delta T$} & 
\colhead{Model\tablenotemark{b}} &
\colhead{$\alpha$} &
\colhead{$\beta$} &
\colhead{$E_\rmn{p}$} &
\colhead{Flux norm.} &
\colhead{$\chi^2$/dof} \\
\colhead{} & 
\colhead{Type} & 
\colhead{(s)} & 
\colhead{(s)} & 
\colhead{} & 
\colhead{} & 
\colhead{} & 
\colhead{(keV)} & 
\colhead{($10^{-6}$~erg~cm$^{-2}$~s$^{-1}$)} & 
\colhead{(Prob.)} 
}
\startdata   
GRB19950210\_T0842 & i,p  & 0.0 & 0.128 & CPL$^\rmn{*}$  & $    -0.01(   -0.48,   +0.62) $ & \nodata               & $    130(   -17,   +23) $ & $    3.7(  -0.5,  +0.5) $ & $  7/17$ (0.978) \\
                   &      &     &       & BAND & $     0.14(   -0.58,   +0.93) $ & $    -3.37(   -6.63,   +0.80) $ & $    123(   -23,   +26) $ & $    4.1(  -0.8,  +1.2) $ & $  7/16$ (0.979) \\
GRB19950211\_T0869 & i,p  & 0.0 & 0.128 & CPL  & $    -0.54(   -0.18,   +0.22) $ & \nodata                         & $    346(   -59,   +78) $ & $   15.5(  -1.9,  +2.1) $ & $ 32/29$ (0.311) \\
                   &      &     &       & BAND$^\rmn{*}$ & $    -0.22( -0.30,  +0.40) $ & $ -2.28( -0.45, +0.23) $ & $    244(   -52,   +75) $ & $   22.0(  -4.4,  +4.8) $ & $ 24/28$ (0.701) \\
GRB19950503\_T6697 & i,p  & 0.0 & 0.128 & PL   & $    -1.30(   -0.05,   +0.05) $ & \nodata  & \nodata                                          & $   77.1( -10.5, +11.0) $ & $ 40/28$ (0.060) \\
                   &      &     &       & CPL$^\rmn{*}$  & $    -1.05(   -0.11,   +0.12) $ & \nodata  & $   2298(  -716, +1413) $              & $   48.4( -10.2, +12.6) $ & $ 21/27$ (0.786) \\
GRB19950520\_T8327 & i,p  & 0.0 & 0.192 & PL   & $    -1.35(   -0.09,   +0.09) $ & \nodata  &  \nodata                                         & $   11.0(  -3.1,  +3.4) $ & $ 36/21$ (0.024) \\
                   &      &     &       & CPL$^\rmn{*}$  & $    -0.46(   -0.42,   +0.56) $ & \nodata  & $    452(  -147,  +358) $              & $    3.2(  -0.9,  +1.5) $ & $ 13/20$ (0.878) \\
                   &      &     &       & BAND & $    -0.40(   -0.44,   +8.62) $ & $    -2.38(   -7.62,   +0.81) $ & $    410(  -292,  +317) $ & $    4.5(  -2.1,  +5.0) $ & $ 12/19$ (0.876) \\
GRB19950805\_T1345 & i,p  & 0.0 & 0.064 & PL   & $    -1.21(   -0.06,   +0.06) $ & \nodata  & \nodata                                          & $   68.0( -13.2, +14.2) $ & $ 31/15$ (0.009) \\
                   &      &     &       & CPL$^\rmn{*}$  & $    -0.84(   -0.16,   +0.21) $ & \nodata  & $   1123(  -424,  +792) $              & $   25.0(  -7.1, +10.1) $ & $ 10/14$ (0.752) \\
                   &      &     &       & BAND & $    -0.51(   -0.39,   +0.71) $ & $    -1.65(   -1.38,   +0.23) $ & $    418(  -240,  +793) $ & $   40.8( -16.4, +16.0) $ & $  7/13$ (0.889) \\
\enddata
\tablenotetext{a}{Relative to the trigger time}
\tablenotetext{b}{The best-fit model is indicated by the asterisk}
\tablecomments{ (This table is available in its entirety in a machine-readable form in the online journal. 
A portion is shown here for guidance regarding its form and content.) }
\end{deluxetable}   
}  
\end{landscape} 
\begin{deluxetable}{crrccccc}
\tabletypesize{\scriptsize}
\tablecaption{Spectral parameters (three-channel spectra)\label{tab:spec_par_3ch}}
\tablewidth{0pt}
\tablehead{
\colhead{Designation} & 
\colhead{$T\rmn{start}$\tablenotemark{a}} &
\colhead{$\Delta T$\tablenotemark{b}} &
\colhead{Model} &
\colhead{$\alpha$} &
\colhead{$E_\rmn{p}$} &
\colhead{Flux} \\
\colhead{} & 
\colhead{(s)} & 
\colhead{(s)} & 
\colhead{} & 
\colhead{} & 
\colhead{(keV)} & 
\colhead{($10^{-6}$ erg~cm$^{-2}$~s$^{-1}$)} 
}
\startdata
 GRB19950414\_T40882 & -0.130  &   0.164 &  CPL & $-0.52(-0.51,+1.03) $ & $  497(-206,+2440) $ &  $4.6(-1.4,+9.8)$ \\ 
 GRB19950526\_T16613 & -0.510  &   1.918 &  CPL & $ 0.32(-0.68,+1.94) $ & $  296(-68,+118)   $ &  $0.9(-0.2,+0.2)$ \\   
 GRB19950610\_T19096 & -0.052  &   0.110 &  CPL & $ 0.72(-0.71,+1.46) $ & $  189(-30,+45)    $ &  $2.7(-0.4,+0.6)$ \\   
 GRB19951013\_T57097 & -0.030  &   0.052 &  CPL & $ 3.96(-3.72,+6.04) $ & $  252(-50,+170)   $ &  $11.0(-1.9,+4.4)$ \\   
 GRB19960312\_T35074 & -0.156  &   0.208 &  CPL & $ 0.24(-0.62,+1.29) $ & $  222(-45,+90)    $ &  $1.8(-0.3,+0.5)$ \\   
\enddata
\tablenotetext{a}{is the burst start time $t_0$}
\tablenotetext{b}{is the burst total duration $T_{100}$}
\tablecomments{ (This table is available in its entirety in a machine-readable form in the online journal. 
A portion is shown here for guidance regarding its form and content.) }
\end{deluxetable}
{\renewcommand\tabcolsep{3pt}
\begin{deluxetable}{ccccccccc}
\tabletypesize{\scriptsize}
\tablecaption{Parameters of CPL+PL model fits\label{tab:extra_comp}}
\tablewidth{0pt}
\tablehead{
\colhead{Designation} & 
\colhead{$\alpha_\rmn{CPL}$} &
\colhead{$E_\rmn{p,CPL}$} &
\colhead{Flux$_\rmn{CPL}$\tablenotemark{a}} &
\colhead{$\alpha_\rmn{PL}$} &
\colhead{Flux$_\rmn{PL}$\tablenotemark{a}} & 
\colhead{$\chi^2/\rmn{dof}$} \\
\colhead{} & 
\colhead{} &
\colhead{(keV)} & 
\colhead{} & 
\colhead{} & 
\colhead{} & 
\colhead{(Prob.)} 
}
\startdata
GRB19960908\_T25028 & $-0.48(-0.47,+0.84)$ & $1528(-282,+357)$  & $27.3(-8.0,+7.0)$   & $-2.07(-0.42,+0.23)$ & $8.7(-5.2,+8.8)$  & 77/63 (0.11) \\
GRB19980205\_T19785 & $-0.70(-0.60,+1.20)$ & $1812(-672,+1333)$ & $13.2(-5.6,+6.2)$   & $-2.22(-0.50,+0.24)$ & $5.4(-3.0,+3.9)$  & 40/55 (0.94) \\
GRB20031214\_T36655 & $-0.31(-0.10,+0.11)$ & $1912(-83,+81)$    & $274.6(-13.4,+12.4)$& $-2.01(-0.39,+0.20)$ & $10.6(-5.6,+8.6)$ & 87/75 (0.15) \\
\enddata
\tablenotetext{a}{In units of $10^{-6}$~erg~cm$^{-2}$}
\end{deluxetable}
} 
\begin{deluxetable}{ccrc}
\tabletypesize{\scriptsize}
\tablecaption{Fluences and Peak Fluxes\label{tab:fl_pf}}
\tablewidth{0pt}
\tablehead{
\colhead{Designation} & 
\colhead{Fluence} &
\colhead{$T_\rmn{pk}$} &
\colhead{Peak Flux} \\
\colhead{} & 
\colhead{($10^{-6}$~erg~cm$^{-2}$)} & 
\colhead{(s)} & 
\colhead{($10^{-5}$~erg~cm$^{-2}$~s$^{-1}$)} 
}
\startdata   
 GRB19950210\_T08424  &  0.63(-0.07,+0.08) & -0.004  &  0.75(-0.18,+0.18) \\
 GRB19950211\_T08697  &  3.26(-0.58,+0.63) &  0.014  &  6.16(-1.37,+1.48) \\
 GRB19950414\_T40882  &  0.76(-0.23,+1.61) & -0.036  &  1.25(-0.49,+2.67) \\
 GRB19950503\_T66971  &  8.03(-1.40,+1.72) &  0.034  & 10.30(-2.57,+3.03) \\
 GRB19950520\_T83271  &  1.42(-0.33,+0.50) &  0.018  &  1.00(-0.38,+0.53) \\
\enddata
\tablecomments{ (This table is available in its entirety in a machine-readable form in the online journal. 
A portion is shown here for guidance regarding its form and content.) }
\end{deluxetable} 
{\begin{landscape}
{\renewcommand\tabcolsep{3pt}
\begin{deluxetable}{crrcrrccccc}
\tabletypesize{\scriptsize}
\tablecaption{Short GRBs with EE\label{tab:EE}}
\tablewidth{0pt}
\tablehead{
\colhead{Designation} & 
\colhead{EE $t_0$\tablenotemark{a}} &
\colhead{EE $T_{100}$} &
\colhead{$T_\rmn{start}$} &
\colhead{$\Delta T$} &
\colhead{Best-fit} &
\colhead{$\alpha$} &
\colhead{$E_\rmn{p}$} &
\colhead{Fluence} &
\colhead{$\chi^2$/dof}\\
\colhead{}    & 
\colhead{(s)} & 
\colhead{(s)} & 
\colhead{(s)} &
\colhead{(s)} &
\colhead{Model} & 
\colhead{} & 
\colhead{(keV)} & 
\colhead{($10^{-6}$~erg~cm$^{-2}$)} & 
\colhead{(Prob.)} 
}
\startdata   
GRB19950503\_T66971  &    6.288  &  109.936  & 0.256  & 78.848  & CPL  & $  -1.61( -0.11, +0.12) $  & $      157(     -24,     +39) $  & $ 41.6(-3.0,+3.9) $  &    100/75 ( 0.03) \\ 
GRB19950526\_T16613  &   28.192  &   64.032  & 41.216  & 23.552  & CPL  & $  -1.15( -0.13, +0.15) $  & $      489(    -125,    +230) $  & $ 13.6(-1.9,+2.6) $  &     74/76 ( 0.54) \\ 
GRB19961225\_T36436  &   14.784  &   12.176  & 8.448  & 24.576  & PL  & $  -1.57( -0.14, +0.16) $  & \nodata  & $ 17.1(-5.6,+7.4) $  &     74/74 ( 0.47) \\ 
GRB19970625\_T23681  &   19.152  &   16.752  & \nodata  & \nodata  & \nodata  & \nodata  & \nodata  & \nodata  & \nodata \\ 
GRB19970923\_T41961  &   27.136  &   29.056  & 24.832  & 32.768  & PL  & $  -1.47( -0.29, +0.30) $  & \nodata  & $  8.3(-5.2,+10.0) $  &     90/65 ( 0.02) \\ 
GRB19980605\_T51131  &    2.160  &  111.760  & \nodata  & \nodata  & \nodata  & \nodata  & \nodata  & \nodata  & \nodata \\ 
GRB19980706\_T57586  &    1.392  &   24.560  & \nodata  & \nodata  & \nodata  & \nodata  & \nodata  & \nodata  & \nodata \\ 
GRB19981107\_T00781  &    2.816  &   16.480  & 8.704  & 24.576  & PL  & $  -1.49( -0.27, +0.28) $  & \nodata  & $  9.5(-4.7,+7.8) $  &     42/72 ( 1.00) \\ 
GRB19981218\_T62134\tablenotemark{b}  &   45.760  &    3.776  & \nodata  & \nodata  & \nodata  & \nodata  & \nodata  & \nodata  & \nodata \\ 
GRB19990313\_T33712  &    2.048  &   70.976  & 8.448  & 65.536  & PL  & $  -1.90( -0.30, +0.41) $  & \nodata  & $  4.4(-2.2,+4.6) $  &     42/63 ( 0.98) \\ 
GRB19990327\_T22911  &    1.136  &   61.328  & 0.256  & 16.384  & CPL  & $  -1.18( -0.19, +0.23) $  & $      389(    -111,    +244) $  & $ 11.2(-1.8,+2.6) $  &     84/62 ( 0.03) \\ 
GRB19990516\_T86065  &    1.408  &   94.016  & 8.448  & 24.576  & PL  & $  -1.85( -0.12, +0.13) $  & \nodata  & $ 18.2(-3.7,+4.5) $  &     52/64 ( 0.86) \\ 
GRB19990712\_T27915  &    6.320  &   34.256  & 8.448  & 32.768  & PL  & $  -2.33( -0.24, +0.28) $  & \nodata  & $  5.2(-1.0,+1.4) $  &     69/63 ( 0.27) \\ 
GRB20000218\_T58744  &    2.400  &   61.344  & 5.888  & 72.960  & PL  & $  -1.60( -0.08, +0.08) $  & \nodata  & $ 80.1(-12.8,+15.3) $  &     71/63 ( 0.23) \\ 
GRB20010317\_T23290  &   25.376  &   27.744  & 25.344  & 8.192  & CPL  & $  -0.29( -0.98, +1.73) $  & $      181(     -46,     +93) $  & $  2.4(-0.6,+0.8) $  &     67/52 ( 0.08) \\ 
GRB20030105\_T52454  &   39.360  &   72.000  & 41.216  & 65.536  & PL  & $  -2.58( -0.47, +0.71) $  & \nodata  & $  2.4(-0.8,+1.3) $  &     46/60 ( 0.91) \\ 
GRB20031214\_T36655  &    2.000  &   70.128  & 8.704  & 65.536  & PL  & $  -1.92( -0.39, +0.46) $  & \nodata  & $  8.2(-3.3,+7.2) $  &     51/61 ( 0.82) \\ 
GRB20040210\_T40272  &    2.016  &    6.032  & \nodata  & \nodata  & \nodata  & \nodata  & \nodata  & \nodata  & \nodata \\ 
GRB20040816\_T29998  &    8.048  &   52.752  & 8.448  & 57.344  & PL  & $  -1.76( -0.17, +0.18) $  & \nodata  & $ 16.4(-5.3,+7.8) $  &     53/60 ( 0.72) \\ 
GRB20050513\_T16804  &    6.048  &    7.520  & \nodata  & \nodata  & \nodata  & \nodata  & \nodata  & \nodata  & \nodata \\ 
GRB20060228\_T44827  &    6.240  &   47.584  & 8.448  & 49.152  & PL  & $  -1.61( -0.14, +0.15) $  & \nodata  & $ 21.4(-6.5,+8.9) $  &     60/59 ( 0.44) \\ 
GRB20061006\_T60326  &    8.960  &  160.768  & 8.448  & 73.728  & PL  & $  -1.38( -0.38, +0.44) $  & \nodata  & $ 19.0(-11.8,+23.5) $  &     59/59 ( 0.46) \\ 
GRB20070915\_T30890  &    2.096  &   54.480  & \nodata  & \nodata  & \nodata  & \nodata  & \nodata  & \nodata  & \nodata \\ 
GRB20071030\_T31964  &   39.744  &   60.608  & \nodata  & \nodata  & \nodata  & \nodata  & \nodata  & \nodata  & \nodata \\ 
GRB20080807\_T85828  &    3.952  &   16.080  & 8.448  & 16.384  & PL  & $  -1.53( -0.13, +0.14) $  & \nodata  & $ 18.9(-4.6,+6.0) $  &     64/65 ( 0.49) \\ 
GRB20090525\_T18274  &    9.296  &   46.768  & 8.448  & 49.152  & PL  & $  -1.72( -0.12, +0.12) $  & \nodata  & $ 15.5(-4.4,+5.8) $  &     55/59 ( 0.63) \\ 
GRB20090720\_T61379  &    2.176  &   14.080  & 0.256  & 8.192  & CPL  & $  -1.30( -0.10, +0.11) $  & $     2250(   -1076,   +2418) $  & $ 16.4(-4.4,+5.1) $  &     94/97 ( 0.56) \\ 
GRB20090831\_T27393  &    3.312  &   80.848  & 0.256  & 40.96  & CPL  & $  -1.42( -0.16, +0.18) $  & $      215(     -48,     +93) $  & $ 14.4(-1.7,+2.2) $  &     66/61 ( 0.31) \\ 
GRB20100714\_T59238  &   10.720  &  137.248  & \nodata  & \nodata  & \nodata  & \nodata  & \nodata  & \nodata  & \nodata \\ 
GRB20100916\_T67270  &    9.520  &   12.768  & 8.448  & 8.192  & PL  & $  -1.58( -0.30, +0.36) $  & \nodata  & $  3.2(-1.7,+3.2) $  &     86/58 ( 0.01) \\ 
\enddata
\tablenotetext{a}{Relative to the trigger time}
\tablenotetext{b}{There is a solar flare in the data at $\sim T_0 + 100$~s}
\end{deluxetable}
}
\end{landscape} 
{\renewcommand\tabcolsep{3pt}
\begin{deluxetable}{ccccccccccccc}
\tabletypesize{\scriptsize}
\tablecaption{Best-fit model parameter distributions\label{tab:pardist}}
\tablewidth{0pt}
\tablehead{
\colhead{Model} &
\colhead{Data} &
\colhead{Number of} &
\multicolumn{2}{c}{$\alpha$} &
\multicolumn{2}{c}{$\beta$} &
\multicolumn{2}{c}{$E_\rmn{p}$ (keV)} &
\multicolumn{2}{c}{Fluence\tablenotemark{b}} &
\multicolumn{2}{c}{Peak Flux\tablenotemark{c}} \\
\colhead{} & 
\colhead{Type\tablenotemark{a}} & 
\colhead{Spectra} & 
\colhead{Median} &
\colhead{CI} &
\colhead{Median} &
\colhead{CI} &
\colhead{Median} &
\colhead{CI} &
\colhead{Median} &
\colhead{CI} &
\colhead{Median} &
\colhead{CI} 
}
\startdata   
      PL &     mult & 4   & $-1.78$ & $[-1.99,-1.61]$ & \nodata & \nodata & \nodata & \nodata & $4.1$ & $[1.4,5.7]$ & $2.1$ & $[0.8,3.0]$  \\
     CPL &     mult & 201 & $-0.47$ & $[-1.14,0.52]$  & \nodata & \nodata & $563$ & $[115,1807]$ & $2.3$ & $[0.5,13.9]$ & $1.5$ & $[0.3,12.8]$  \\
    BAND &     mult & 9   & $-0.12$ & $[-1.13,1.42]$  & $-2.28$ & $[-3.15,-1.74]$ & $204$ & $[40,364]$ & $3.8$ & $[1.9,39.1]$ & $2.8$ & $[0.5,7.3]$  \\
     all &     mult & 214 & \nodata & \nodata & \nodata  & \nodata & \nodata & \nodata & $2.4$ & $[0.5,20.1]$ & $1.6$ & $[0.4,12.8]$ \\
     CPL &     3ch  & 79  & $-0.36$ & $[-1.23,0.90]$ & \nodata & \nodata & $459$ & $[190,1180]$ & $0.9$ & $[0.3,3.4]$ & $1.0$ & $[0.4,4.0]$  \\              
\enddata
\tablenotetext{a}{Multichannel spectrum~--- ``mult'' or three-channel spectrum~--- ``3ch''}
\tablenotetext{b}{In units of $10^{-6}$~erg~cm$^{-2}$}
\tablenotetext{c}{In units of $10^{-5}$~erg~cm$^{-2}$~s$^{-1}$}
\end{deluxetable}
} 

\end{document}